\documentclass[12pt]{article}
\usepackage[margin=0.7in]{geometry}
\usepackage{float}
\usepackage{amsmath}
\usepackage{amssymb}
\usepackage{graphicx}
\usepackage{cite}
\usepackage{color}
\usepackage{dcolumn}
\usepackage{bm}
\usepackage{booktabs}
\usepackage{newfloat}
\DeclareFloatingEnvironment[name={SI Figure}]{suppfigure}
\DeclareFloatingEnvironment[name={SI equation}]{suppequation}
\DeclareFloatingEnvironment[name={SI Table}]{supptable}
\usepackage{hyperref}
\begin{document}

\begin{flushleft}
{\LARGE
\textbf{Spectral properties of complex networks}
}
\\
Camellia Sarkar$^{1}$ \& Sarika Jalan$^{1,2,3,\ast}$
\\ 
\it ${^1}$ Centre for Biosciences and Biomedical Engineering, Indian Institute of Technology Indore, Khandwa Road, Simrol, Indore, India 453552\\
\it ${^2}$ Complex Systems Lab, Discipline of Physics, Indian Institute of Technology Indore, Khandwa Road, Simrol, Indore, India 453552\\
\it ${^3}$ Lobachevsky University, Nizhny Novgorod, Gagarin avenue 23, Russia 603950\\

$\ast$ Corresponding author e-mail: sarikajalan9@gmail.com
\end{flushleft}

\begin{abstract}
This review presents an account of the major works done on spectra of adjacency matrices drawn on networks and the basic understanding attained so far. We have divided the review under three sections: (a) extremal eigenvalues, (b) bulk part of the spectrum and (c) degenerate eigenvalues, based on the intrinsic properties of eigenvalues and the phenomena they capture. We have reviewed the works done for spectra of various popular model networks, such as the Erd\H{o}s-R\'enyi random networks, scale-free networks, 1-d lattice, small-world networks, and various different real-world networks. Additionally, potential applications of spectral properties for natural processes have been reviewed.
\end{abstract}

{\bf Various natural and man-made systems have been modeled under the network theory framework. Different network models with distinct design principles have been proposed to better understand these real-world networks. The eigenvalue spectrum of these networks not only contain information about
structural characteristics of underlying networks but also provide insight to dynamical behaviour and stability of
corresponding complex systems. Depending on the structural characteristics of underlying model networks, the spectra of these networks exhibit specific features. All these ascertain that the spectra of networks can be used as a practical tool for classifying and understanding different real-world systems represented as networks. In this review, we first discussed the features which the different regions of spectra furnish, in case of the model networks. Further, we went on to discuss spectral characteristics of real-world networks, with particular emphasis on their extent of similarities and differences with the spectra of model networks.}

\section{Introduction}
Networks present a simple framework to model complex systems comprising of interacting elements. The two basic ingredients of a network are its nodes and connections. Mathematically, a network or a graph is defined as a set of $N$ nodes and $N_c$ connections
which can be represented in terms of an adjacency matrix, {\bf A} as:
\begin{equation}
A_{\mathrm {ij}} = \begin{cases} 1~~\mbox{if } i \sim j \\
0 ~~ \mbox{otherwise} \end{cases}
\end{equation} 
It is useful to define here the degree of a node, the largest degree and the 
average degree of a network. Degree of
a node $i$ refers to the number of nodes $i$ is connected to and can be calculated from
Eq.~1 as $k_i = \sum_{j=1}^{N} A_{ij}$. Similarly the largest degree ($k_{max}$) is {\it max}$(k_{i})$ where $1 \leq i \leq N$, and
the average degree $\langle k \rangle = \frac{\sum_{i=1}^{N} k_{i}}{N}$. 

In the past two decades, network science has shown tremendous success in modeling and understanding complex phenomena in various model and real-world systems. Research in network science has elucidated the importance of interactions which has led to fundamental understanding of emergent phenomena in various complex systems and processes, for instance, cellular signalling, disease spread, scientific collaboration, transportation, WWW, power grid and so on \cite{Barabasi_rev_2002,Boccaletti_Rev}. Many of these networks appear to share certain nontrivial, similar patterns in connections between their elements. An understanding to
the origins of these patterns and identifying and characterizing new ones is one of the main driving forces for research in complex networks. Apart from various investigations which focus on direct measurements of the structural properties of
networks, there have been studies demonstrating that
properties of networks or graphs could be well characterized
by the spectrum of associated adjacency matrix \cite{Cvetkovic}. The spectrum of a network is the set of eigenvalues of its adjacency matrix ($A_{ij}$) and is denoted as $\lambda_i$, where $i=1,2, \hdots, N$ such that $\lambda_1>\lambda_2 \geq \lambda_3 \geq \hdots \geq \lambda_N$.
For an undirected network, the adjacency matrix is symmetric and consequently has real eigenvalues. For a directed network, the adjacency matrix is asymmetric and has complex eigenvalues. Further, there can be networks having weighted connections, negative couplings, etc. Note that here we have restricted ourselves to symmetric matrices generated from networks, having binary interactions. Further, we consider simple
graphs, i.e., graphs devoid of self and multiple connections. 

Spectra is shown to provide information
about the basic topological properties of the underlying
networks \cite{Cvetkovic,Spectra_book_Doob} as well as have
been used to understand dynamical properties of
interacting chaotic units on networks \cite{Restrepo_PRE_2005,Restrepo_PRL_2006}. For instance, the largest eigenvalue, one of the extremal eigenvalues of a network has been used to understand dynamics of epidemic spread in the corresponding system \cite{Meighem_2009}. The spectral density has been shown to capture various universality features existing in real-world systems \cite{Farkas_2001}. Furthermore, degeneracy at $0$ eigenvalue in biological networks has been known to be associated with the gene duplication phenomenon \cite{Kamp_PRE_2005}. These are few of the examples to motivate the readers about the importance of network spectra as well as to justify the need of segregating this review into three broad divisions, namely (i) extremal eigenvalues, (ii) bulk of the spectra and (iii) degenerate eigenvalues. 
The extremal eigenvalues are referred to those eigenvalues which are far
separated from the rest of the eigenvalues (referred as the bulk part). The degenerate eigenvalues
are also part of the bulk part, however, degenerate eigenvalues provide clues to
specific symmetrical patterns in the underlying networks and hence have been segregated in
a different section.

\section{Extremal eigenvalues}
The largest eigenvalue of the network adjacency matrix (denoted as $\lambda_{1}$) has emerged as a key quantity important for the
study of a variety of systems and processes, such as virus spread, synchronization \cite{kurths_syn} of coupled oscillators, stability of couplings in brain, etc.
For example, it was shown that the inverse of the largest eigenvalue equals the epidemic threshold for virus propagation \cite{Chakrabarti_2003,Chakrabarti_2008}. Epidemic threshold can be understood as the critical number or density of susceptible hosts required for an epidemic to occur. A null epidemic threshold was reported in case of scale-free networks, rendering scale-free networks prone to the spreading and the persistence of infections at
whatever spreading rate the epidemic agents possess \cite{Vespignani_PRL_2001,Vespignani_PRL_2003,Castellano_PRL_2010}. Further, in the case of networks of coupled phase oscillators, the critical coupling strength (say $\tau_{c}$) at which the transition from incoherence to coherence occurs is determined by the largest eigenvalue of the adjacency matrix \cite{Restrepo_PRE_2005} as $\tau_{c} \propto \frac{1}{\lambda_{1}}$. 
Further, the largest eigenvalues of network's adjacency matrices are known to determine the stability of various
interaction patterns \cite{stability_largest_eigen1,Restrepo_PRE_2005,stability_largest_eigen3,stability_largest_eigen4}.
Considering the crucial information that $\lambda_{1}$ possesses and its importance, it would be 
indeed important to explore properties of the largest eigenvalue and its behaviour in different systems and networks. In the following, first we will discuss the bounds of eigenvalues for adjacency matrices with non-negative entries followed by the largest eigenvalue behaviour for popular model networks, such as Erd\H{o}s-R\'enyi (ER) random network, scale-free network, small-world network and 1-d lattice. Finally,
we include a discussion on the largest eigenvalue for networks constructed from various real-world data.

As per the Perron-Frobenius theorem, the largest eigenvalue of a matrix with non-negative entries is real and positive. According to the Gerschgorin theorem, {\it every eigenvalue of an adjacency matrix ($A$) lies in at least one of the circular discs with centre $a_{ii}$ and radii [$\sum_{j=1; j \neq i} |a_{ij}|$, $\sum_{j=1;j \neq i} |a_{ji}|$]} and can be given as \cite{Meighem_book_2010}
\begin{equation}
|a_{ii} - \lambda| \leq \sum_{j=1; j \neq i}^{N} |a_{ij}| 
\end{equation}
For the simple graphs considered here, $a_{ii}=0$ and therefore the largest radius of the largest
circle is $k_{max}$, where $k$ denotes degree of a node and $k_{max}$ refers to the maximum degree. 
Consequently, all the eigenvalues of an adjacency matrix lie in the 
interval [$-k_{max},k_{max}$]. Furthermore, as per as another theorem \cite{Cvetkovic},
the largest eigenvalue of the adjacency matrix of networks are bound by the average degree ($\langle k \rangle$) and largest degree ($k_{max}$) of the corresponding network as  
\begin{equation}
\langle k \rangle \leq \lambda_{1} \leq k_{max}
\label{lambda_max_bound}
\end{equation} 
For a globally connected network, the largest eigenvalue $\lambda_{1} = (N-1)$ \cite{Meighem_book_2010}. 
\begin{figure}
\centerline{\includegraphics[width=0.99\columnwidth]{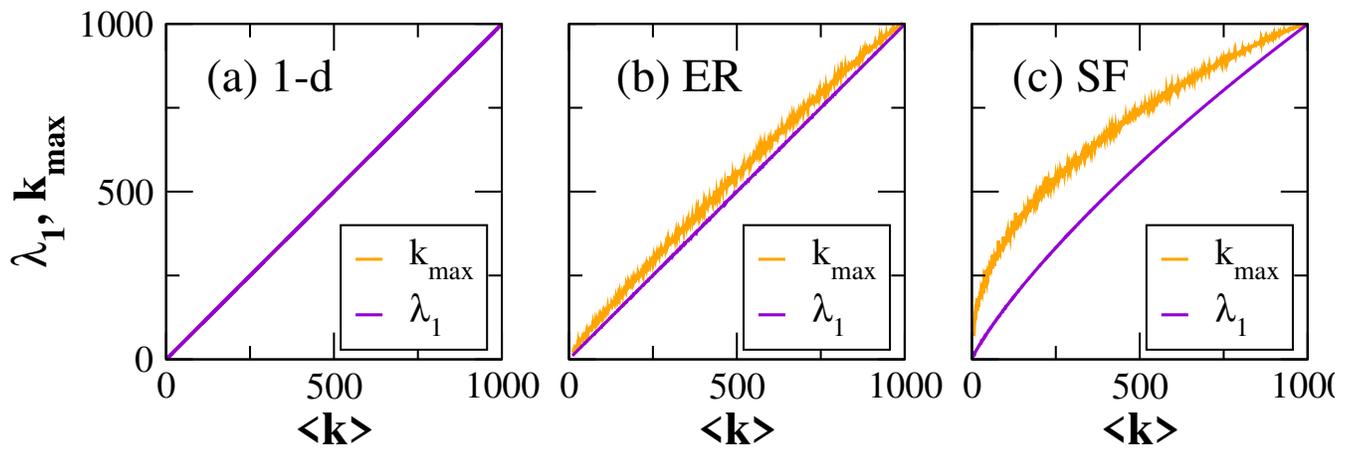}}
\caption{Plots of the largest eigenvalue ($\lambda_{1}$) and the largest degree ($k_{max}$) as a function of average degree ($\langle k \rangle$) for (a) 1-d lattice, (b) ER random network and (c) scale-free network, all having network size, $N = 1000$.}
\label{avg_deg_larg_eig}
\end{figure}

Depending on various underlying structural properties of networks, the largest eigenvalue may have different relations with the largest degree. In the following, we discuss the construction of the various model networks, their structural signatures and also focus on how the largest eigenvalue and its 
fluctuations around the average value varies with change in different structural properties
for an ensemble of networks.
%The various model networks, such as the Erd\"os-Ren\'yi (ER) random network, scale-free network, small-world network and 1-d lattice have different structural properties.

A 1-d lattice (also known as regular network or ring lattice) refers to a ring of $N$ nodes, 
each connected to its $k$ nearest neighbours by undirected edges. In this regular network, 
each node has exactly the same degree and the graph is always connected. Hence $\lambda_{1} = \langle k \rangle = k_{max}$. Since $\lambda_{1}$ and $k_{max}$ are exactly same in case of 1-d lattice, these two quantities exhibit linear increase with increase in the average degree (Fig~\ref{avg_deg_larg_eig} (a)). 

Starting with $N$ nodes, ER random networks are constructed by connecting every pair of nodes with a probability, $p$. An ER random network has approximately $pN(N-1)/2$ connections distributed randomly.
%The majority of nodes in this network have their degree close to the average degree $\langle k \rangle$ of the network, given as $\langle k \rangle = p(N-1) \simeq pN$. 
The degree distribution ($P(k)$) of an ER random graph follows a binomial distribution, with the expected value of the degree of all the nodes being equal to $\langle k \rangle = p(N-1) \simeq pN$. %The largest eigenvalue of an ER random network scales as $pN \approx \langle k \rangle$ \cite{Farkas_2001}.
The largest eigenvalue of an ER random network takes a value
\begin{equation}
\lambda_{1} \approx [1 + o(1)] \langle k \rangle
\label{Bound_ER}
\end{equation}
provided $pN >> log(N)$ \cite{Chung_PNAS_2003}. With an increase in $\langle k \rangle$, the largest degree takes a higher value as compared to the largest eigenvalue in case of ER random networks (Fig~\ref{avg_deg_larg_eig} (b)). This can be understood by combining Eqs.~\ref{lambda_max_bound} and \ref{Bound_ER}
that in ER random networks, the degrees of all the vertices have the same expected value, i.e. $Np$.

Another popular model network is the scale-free network, whose discovery by Barab\'asi and Albert in 1999 marked the rebirth of network science. According to the Barab\'asi-Albert (BA) model \cite{Barabasi_Science_1999}, the scale-free networks are generated based on two mechanisms: growth and preferential attachment, and is popularly known as `rich-gets-richer' model. 
In the BA model, starting with a small number of nodes, in each time step a 
new node is added with certain number of connections. This newly added node preferentially connects with a already existing high degree node $i$ with probability $\pi(k_{i}) = \frac{k_{i}}{\sum_{j}k_{j}}$.
The scale-free networks are characterized by degree distribution following power-law ($P(k) \sim k^{-\gamma}$) \cite{Barabasi_rev_2002}, where $\gamma$ refers to the power-law exponent. A wide range of real-world networks have been shown to exhibit power-law behaviour, with $\gamma$ lying in the range $2 < \gamma < 3$. For different values of $\gamma$, the largest eigenvalue of the scale-free networks exhibits different bounds. The largest eigenvalue of the adjacency matrix of a scale-free network depends on the largest degree ($k_{max}$) and the second moment of the degree distribution ($\frac{\langle k^{2} \rangle}{\langle k \rangle}$), given as \cite{Chung_PNAS_2003,Chung_2003}
\begin{equation}
\lambda_{1} = \begin{cases} (1+o(1)) \sqrt{k_{max}},~~\mbox{if } \sqrt{k_{max}} > \frac{\langle k^{2} \rangle}{\langle k \rangle} ln^{2}(N)\\
(1+o(1)) \frac{\langle k^{2} \rangle}{\langle k \rangle}, ~~~~~ \mbox{if} \frac{\langle k^{2} \rangle}{\langle k \rangle} > \sqrt{k_{max}} ln(N) \end{cases}
\label{SF_lambda}
\end{equation}
where $\frac{\langle k^{2} \rangle}{\langle k \rangle} = \frac{\sum_{i=1}^{N} k_{i}^2}{\sum_{i=1}^{N} k_{i}}$ \cite{Chung_2003}. In the case of scale-free networks, the largest degree is a growing function of $N$, which, for uncorrelated networks \cite{Dorogovtsev_2002} takes the value $k_{max} \sim N^{\frac{1}{2}}$ for $\gamma \leq 3$ and $k_{max} \sim N^{\frac{1}{\gamma - 1}}$ for $\gamma >3$ \cite{Vespignani_EPJB_2004}. For very large networks, the algebraic increase in $k_{max}$ allows us to disregard the logarithmic terms of Eq.~\ref{SF_lambda}, thus leading to a simpler expression that holds good for any value of $\gamma$ \cite{Castellano_PRL_2010}
\begin{equation}
\lambda_{1} \approx max[\sqrt{k_{max}}, \frac{\langle k^{2} \rangle}{\langle k \rangle}]
\label{SF_lambda_2}
\end{equation}
Fig~\ref{avg_deg_larg_eig} (c) depicts the behaviour of $\lambda_{1}$ and $k_{max}$ with increasing value of $\langle k \rangle$. From Eq.~\ref{SF_lambda_2}, it follows that $\lambda_{1}$ exhibits values lower than $k_{max}$ for all values of $\langle k \rangle$. For $\gamma > 3$, the ratio of the moments is finite and 
it is clear that the largest eigenvalue is governed by $k_{max}$. This also remains true for $2.5 < \gamma < 3$, since in that range $\frac{\langle k^{2} \rangle}{\langle k \rangle} \sim k_{max}^{3 - \gamma} \ll \sqrt{k_{max}}$ \cite{Castellano_PRL_2010}. Only for $2 < \gamma < 2.5$ the largest eigenvalue is set by the moments of the degree distribution. Imposing this condition on Eq.~\ref{SF_lambda_2}, it is possible to obtain a simpler expression on the bound of the largest eigenvalue for different ranges of $\gamma$, given as
\begin{equation}
\lambda_{1} \approx \begin{cases} \sqrt{k_{max}}~~\mbox{if } \gamma > 2.5 \\
\frac{\langle k^{2} \rangle}{\langle k \rangle} ~~ \mbox{if } 2 < \gamma < 2.5\end{cases}
\label{SF_lambda_3}
\end{equation}
The implication and importance of this bound (Eq.~\ref{SF_lambda_3}) has been realised while determining the epidemic threshold in scale-free networks \cite{Restrepo_PRE_2005}. Since $k_{max}$ grows as a function of $N$ for any $\gamma$, as per Eq.~\ref{SF_lambda_3} and Fig.~\ref{avg_deg_larg_eig} (c), the threshold for epidemic spread (characterized as the inverse of the largest eigenvalue) goes to zero as the system size goes to infinity \cite{Castellano_PRL_2010}.
\begin{figure}
\centerline{\includegraphics[width=0.9\columnwidth]{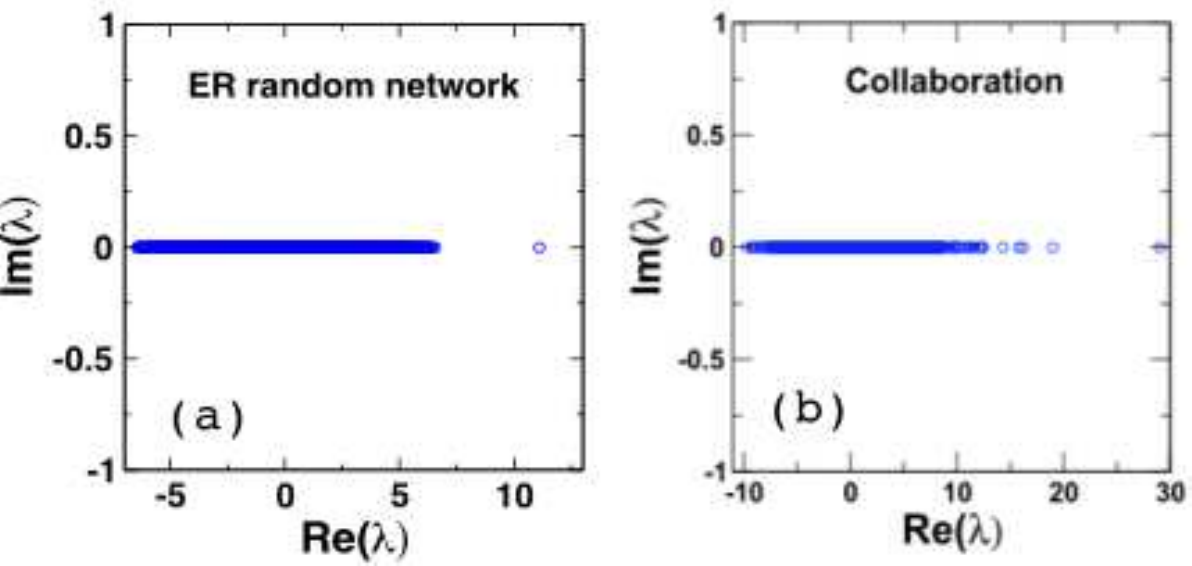}}
\caption{Eigenvalues of undirected (a) ER random network and (b) scientific collaboration network, plotted on real and imaginary axes. Eigenvalues separated from the bulk of the eigenvalues indicate existence of communities, depicted in case of scientific collaboration network. $\textcopyright$ 2016 IEEE. Reprinted figure (Figure 4 left panel), with permission, from C. Sarkar and S. Jalan, IEEE Trans. Comput. Social Syst. 3 (3), 132 (2016). DOI: \href{https://doi.org/10.1109/TCSS.2016.2591778}{10.1109/TCSS.2016.2591778}.}
\label{Eig_sep_bulk}
\end{figure}
\begin{figure}
\centerline{\includegraphics[width=0.99\columnwidth]{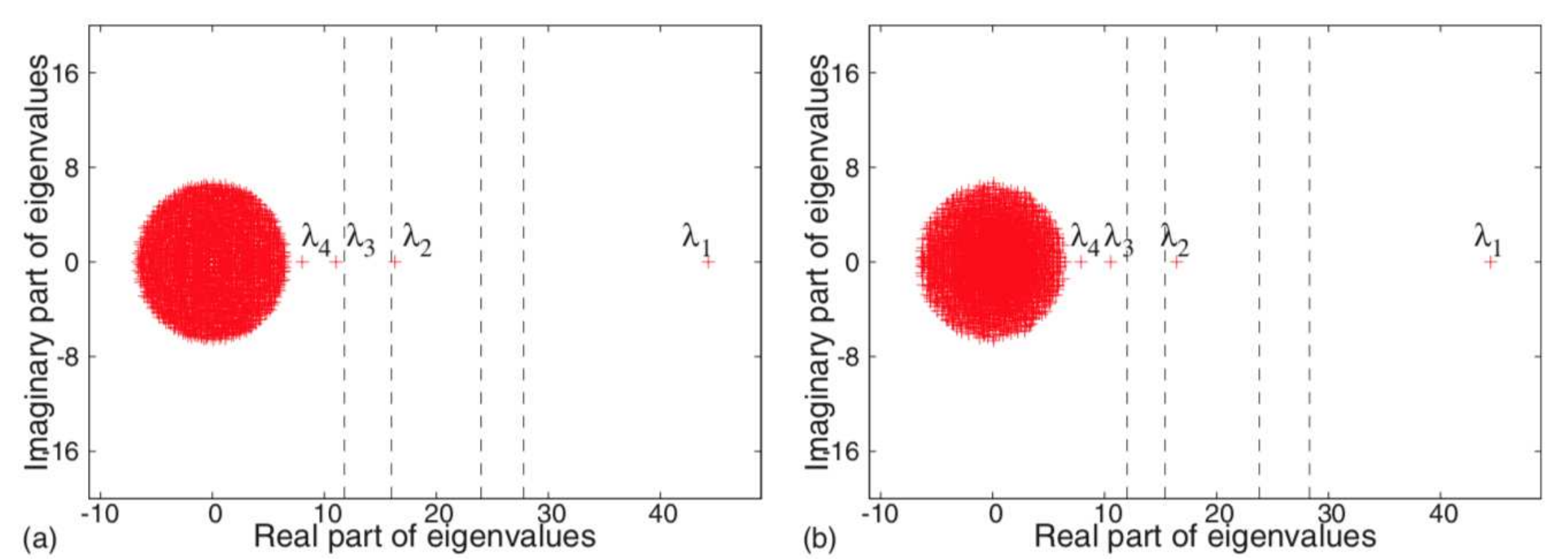}}
\caption{Plots of real and imaginary parts of eigenvalues of directed (a) ER network and (b) scale-free network having four unequal-sized communities each. The average number of within community and between community links are equal in the both the networks. Four eigenvalues lie outside the cloud of rest of the eigenvalues, corresponding to the four communities present in the network. Reprinted figure (Figure 4) with permission from
S. Chauhan, M. Girvan, and E. Ott, Phys. Rev. E 80, 056114 (2009). Copyright (2009) by the
American Physical Society. DOI: \href{https://doi.org/10.1103/PhysRevE.80.056114}{10.1103/PhysRevE.80.056114}.}
\label{Comm_ER_SF}
\end{figure}

Until now it is clear that the largest eigenvalue of an adjacency matrix is largely governed by the largest degree of the underlying network. For some other works, it has been shown that other 
structural parameters of the corresponding networks are related to the largest eigenvalue. For instance, the degree-degree correlations (denoted as $r$) exhibit profound impact on the largest eigenvalue. Analytically it has been shown that positive (assortative) degree-degree correlations ($r > 0$) leads to increase in $\lambda_1$, while negative (disassortative) degree-degree correlations ($r < 0$) decrease $\lambda_1$ \cite{Meighem_EPJB_2010,Dorogovtsev_PRL_2012}. Further, in networks having community structure, the largest eigenvalue has been known to be associated with the largest degree of the subnetwork containing the largest degree node of the network \cite{Castellano_PRX_2017}. Other networks having correlated and uncorrelated topology or having a given degree distribution have shown different bounds for the largest eigenvalues \cite{Chung_PNAS_2003,Chung_2003}.

Further, for an undirected ER random network, only one eigenvalue, i.e. the largest eigenvalue is situated away from the rest of the eigenvalues (Fig.~\ref{Eig_sep_bulk}(a)).
A set of eigenvalues separated from the bulk of the eigenvalues indicate the existence of 
communities in the underlying network. For example, in the case of network of scientific collaborators, few eigenvalues are separated from the bulk of the eigenvalues (clearly seen in Fig.~\ref{Eig_sep_bulk}(b)), indicating presence of collaboration modules among scientists \cite{SJ_IEEE_2016}. For a network with $N$ nodes and $m$ communities, there will typically be $m$ eigenvalues that are significantly larger than the magnitudes of all the other ($N$-$m$) eigenvalues \cite{Chauhan_Ott}. Temporal multilayer network analysis of portfolio correlations of bank-firm credit market of Japan over a period of 22 years across different layers (total loans, short-term loans and long-term loans) revealed that the largest eigenvalue is always separated from the rest of the eigenvalues \cite{Bank_Japan_eig}. The largest eigenvalues were found to contribute the most to the total variance, measured in terms of absorption ratio given as $E_{i} = \frac{\sum_{j=1}^{i}\lambda_{j}}{N}$ for $i = 1, 2, .... N$. By relating the pattern of absorption ratios for different eigenvalues across 22 years with the events of financial crisis in Japan, it was demonstrated how the largest eigenvalue acts as an indicator of higher level of systemic risk in the market \cite{Bank_Japan_eig}. The relation between eigenvalues separated from the bulk and the corresponding number of communities existing in the underlying network structure has been further verified numerically by analyzing the eigenvalues of networks having known number of communities \cite{Chauhan_Ott}. It has been shown that in case of directed ER random networks and scale-free networks, the number of communities they have, equals the exact number of eigenvalues that lie outside the bulk of the eigenvalues (Fig.~\ref{Comm_ER_SF}), thus ascertaining the claim.
\begin{figure}
\centerline{\includegraphics[width=0.75\columnwidth]{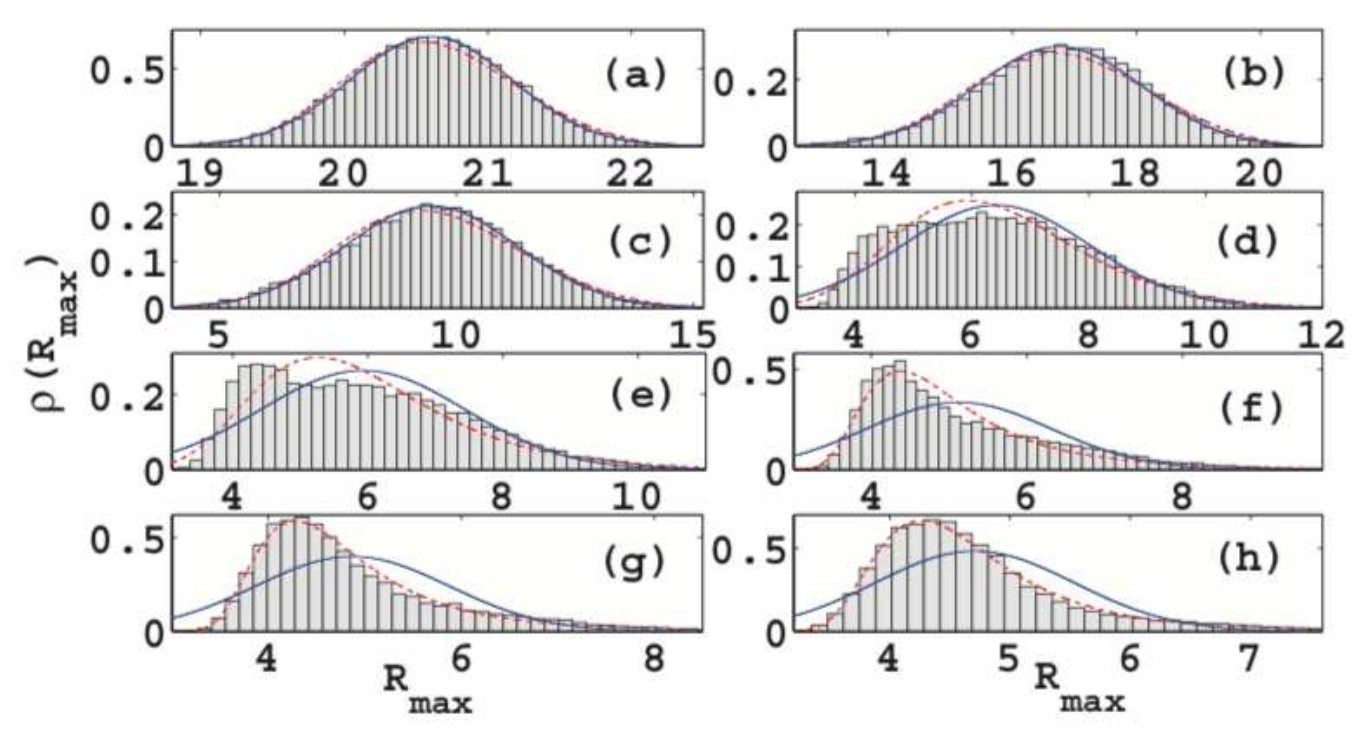}}
\caption{Distribution of largest real part of eigenvalues for ER random networks with $N=100$ and $\langle k \rangle = 20$ for various inhibitory probabilities ($p_{in}$). Panels (a), (b), (c), (d), (e), (f), (g), and (h) correspond to $p_{in}$ = 0, 0.1, 0.3, 0.4, 0.42, 0.46, 0.48, and 0.5, respectively. The histograms represent numerical result, and the solid and dashed lines are obtained by fitting the data with the normal and the GEV distributions, respectively. This figure depicts transition of the largest eigenvalue statistics from normal to GEV distribution with change in $p_{in}$. Reprinted figure (Figure 4) with permission from
S. K. Dwivedi and S. Jalan, Phys. Rev. E 87, 042714 (2013). Copyright (2013) by the
American Physical Society. DOI: \href{https://doi.org/10.1103/PhysRevE.87.042714}{10.1103/PhysRevE.87.042714}.}
\label{Extreme_ER}
\end{figure}

\begin{figure}
\centerline{\includegraphics[width=0.6\columnwidth]{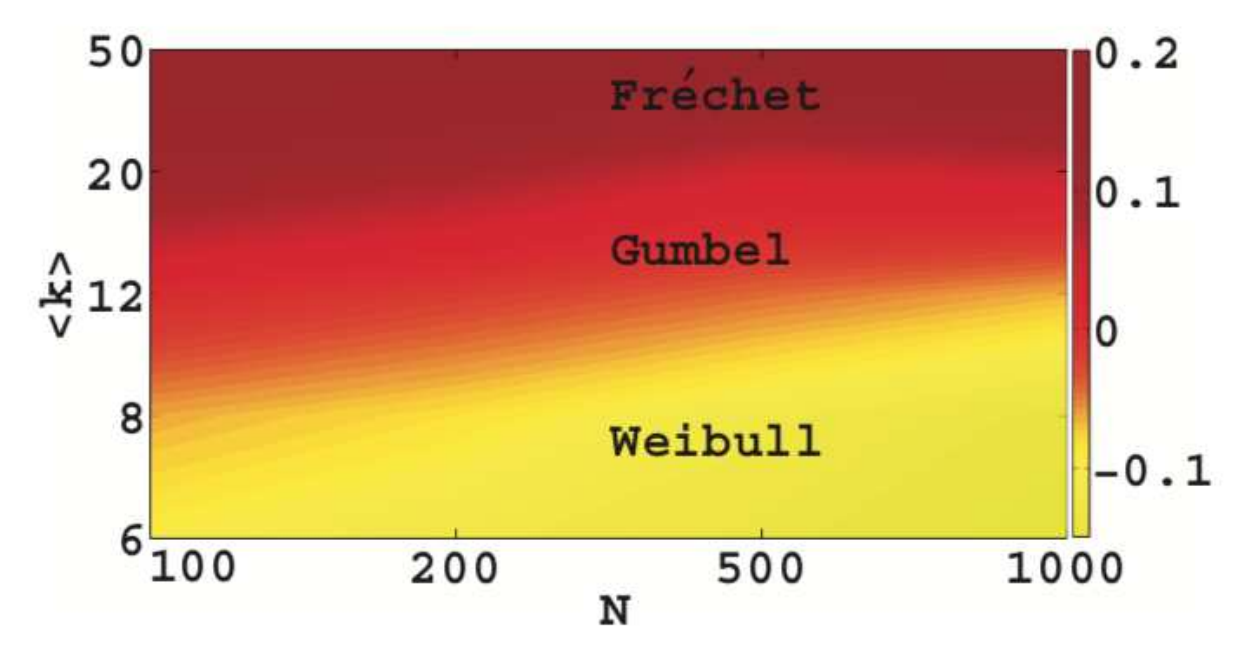}}
\caption{Phase diagram in two-parameter space, $N$ and $\langle k \rangle$, elucidating the nature of the GEV statistics based on the value of the shape parameter $\xi$ and the tail of the distribution at $p_{in} = 0.5$. Reprinted figure (Figure 5) with permission from
S. K. Dwivedi and S. Jalan, Phys. Rev. E 87, 042714 (2013). Copyright (2013) by the
American Physical Society. DOI: \href{https://doi.org/10.1103/PhysRevE.87.042714}{10.1103/PhysRevE.87.042714}.}
\label{GEV_phase}
\end{figure}

Along with the bounds of the largest eigenvalue, there are works which have tried to understand the statistics of the largest eigenvalue. Various approximations, such as mean-field approximation, linear approximation and Markov model have been used to better understand the largest eigenvalue statistics \cite{Restrepo_2007}. It has been found that though the ER random networks and scale-free networks have different bounds of the largest eigenvalue, they yield similar statistics, i.e. normal distribution \cite{SJ_PRE_2013}. Although here in this review so far we have focused on networks having symmetric matrices, there exist networks having directed architecture. For example, in ecological systems, predator-prey interactions yield networks having asymmetric matrices. Robert May, in his celebrated work drawn on ecological systems demonstrated that the largest real part of an eigenvalue of a corresponding adjacency matrix contains information about the stability of the underlying system \cite{May_1972}.   
In neuroscience, networks of neurons are often studied using
models in which interconnections are represented by a synaptic
matrix with elements drawn randomly \cite{stability_largest_eigen3,Cessac_2006}. Eigenvalues of
these matrices are useful for studying spontaneous activities
and evoked responses in such models \cite{stability_largest_eigen3,Vogels_2005}, and the existence
of spontaneous activity depends on whether the real part of
any eigenvalue is large enough to destabilize the silent state
in a linear analysis. Motivated largely by neuroscience and ecological systems, there 
exist models incorporating both the positive and negative types of couplings in the underlying networks.
The presence of negative couplings in a system yields richer statistics of eigenvalues. For inhibitory couplings introduced with a probability $p_{in}$, the entries of adjacency matrices can be written as,
\begin{equation}
A_{\mathrm {ij}} = \begin{cases} 1,~~~~\mbox{if } p_{in} = 0 \\
-1, ~~\mbox{if } p_{in} > 0 \\
0, ~~~~ \mbox{otherwise} \end{cases}
\end{equation} 
The largest eigenvalue of this matrix also follow the same bound as given by Eq.~2. The only difference is that that due to inhibitory couplings, the matrix is no more symmetric and the eigenvalues lie on
the complex plane. It has been shown for such networks containing both the positive and negative 
couplings, the real part of the largest eigenvalue ($R_{max}$)
statistics may follow generalized extreme value (GEV) statistics, depending upon the ratio of positive-negative couplings as well as the structural properties of the underlying networks. The probability density function for extreme value statistics is given by \cite{Gumbel}
\begin{equation}
\rho(x) = \begin{cases} \frac{1}{\sigma}[1+{(\xi \frac{x - \mu}{\sigma})}]^{-1-\frac{1}{\xi}} exp[-(1+{(\xi \frac{x - \mu}{\sigma})}^{-\frac{1}{\xi}})], ~~ \mbox{if } \xi \neq 0 \\
\frac{1}{\sigma} exp(-\frac{x-\mu}{\sigma}) exp[-exp(-\frac{x-\mu}{\sigma})], ~~ \mbox{if } \xi = 0 \end{cases}
\label{GEV}
\end{equation} 
where $\mu$, $\sigma$ and $\xi$ denote location parameter, scale parameter and shape parameter, respectively.
Distributions associated with $\xi > 0, = 0,$ and $< 0$ are characterized by Fr\'echet, Gumbel, and Weibull distributions, respectively. At a certain ratio of positive and negative couplings, the largest eigenvalue statistics manifests a transition to GEV distribution from Eq.~\ref{GEV} (Fig.~\ref{Extreme_ER}). For lower values of $p_{in}$, the nature of the distribution is normal (Fig.~\ref{Extreme_ER} (a)-(d)), which is followed by a regime of $p_{in}$ where the distribution can be modeled using extreme value statistics. Fig.~\ref{Extreme_ER} (e)-(h) indicates that the largest eigenvalue statistics converges to Weibull distribution of GEV. The transition to GEV statistics is further governed by $N$ and $\langle k \rangle$ of the underlying networks. For small $\langle k \rangle$, the GEV statistics correlated to Weibull distribution, whereas with an increase in connection probability it indicates a transition to Fr\'echet distribution through Gumbel (Fig.~\ref{GEV_phase}).
The statistical properties of the largest real part of eigenvalues of synaptic matrices capturing positive and negative couplings reveal a transition to the extreme-value distribution, which has enabled in attaining fundamental understanding pertaining to the stability of the underlying systems as well as capturing extreme events. 

In a nutshell, this section provides us an understanding of the importance of the largest eigenvalue by accounting various works done in the context of extremal eigenvalues, highlighting their bounds and the statistics they yield. In the next section, we will discuss the properties of complex networks that are captured by the bulk part of the spectrum.

\section{Bulk of the eigenvalues}
The largest eigenvalue provides insight into the stability and dynamics of and on networks, and hence is very important for large class of applications. However, it was further realized that
the rest part of the spectrum is useful to understand randomness in interactions as well as can be used to assign universality class to underlying networks. 
The largest eigenvalue resides away from the rest of the eigenvalues which form the bulk part of the spectrum. The bulk of the eigenvalues form a circular cloud entered approximately at the origin \cite{Chauhan_Ott}. The root-mean-square radius of the cloud has an upper bound given by $\sqrt{\langle k \rangle}$. In the following, we will discuss how the bulk of the eigenvalues are distributed in case of different model networks as well as in networks constructed from real data. We will also discuss about the spacings among eigenvalues and the rich statistics they yield in case of model and real-world networks, captured through the random matrix theory (RMT). 
\begin{figure}
  \centerline{\includegraphics[width=0.5\columnwidth]{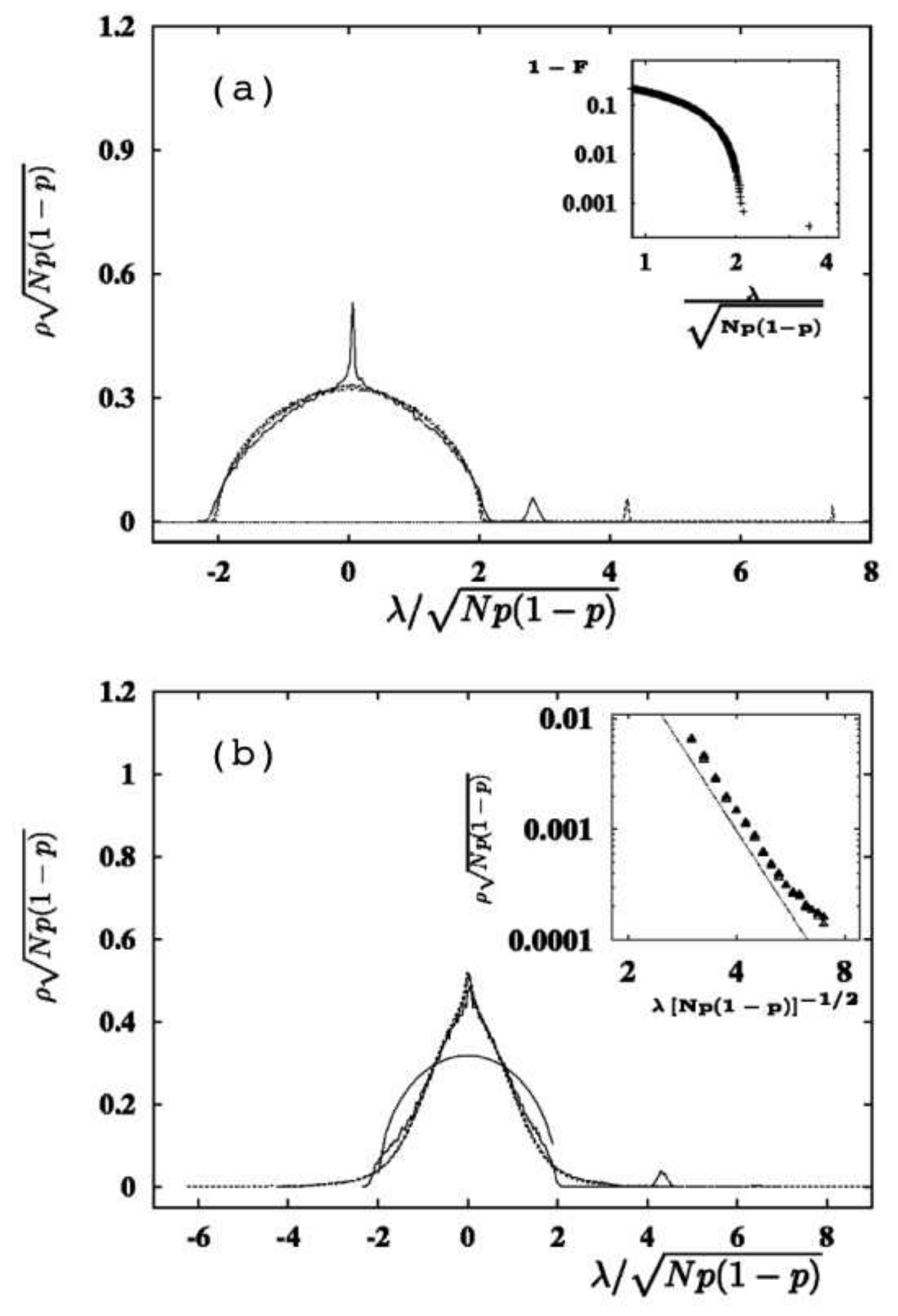}}
  \caption{Spectral density of the adjacency matrix of (a) ER random network ($N=3000, \langle k \rangle=5$) and (b) scale-free network ($N=40000, \langle k \rangle=10$). The spectral density of
an ER random network converges to a semicircle. {\it Inset of (a):} At the edge of the semicircle, the spectral density decays exponentially. The spectral density of a scale-free network exhibits a triangle-like structure. {\it Inset of (b):} The edges of spectral density of the scale-free network plotted on a doubly logarithmic scale show a power-law decay. Reprinted figures (Figures 1 and 4) with permission from
I. J. Farkas, I. Der\'enyi, A.-L. Barab\'asi, and T. Vicsek, Phys. Rev. E, 64 (2), 026704 (2001). Copyright (2001) by the
American Physical Society. DOI: \href{https://journals.aps.org/pre/abstract/10.1103/PhysRevE.64.026704}{10.1103/PhysRevE.64.026704}.}
  \label{Spectral_density_ER}
\end{figure}
\subsection{Spectral density}
The spectral density of a graph is the density of the eigenvalues of its adjacency matrix. For a finite system, this can be written as a sum of $\delta$ functions as
\begin{equation}
\rho(\lambda) = \frac{1}{N} \sum_{j=1}^{N} \delta(\lambda - \lambda_{j}), 
\end{equation} 
which converges to a continuous function with $N \rightarrow \infty$ ($\lambda_{j}$ is
the $j^{th}$ eigenvalue of the graph's adjacency matrix, when the eigenvalues are sorted in descending order).
If $A$ is a real symmetric
$N \times N$ random matrix, it was found that in the limit $N \rightarrow \infty$, the density of the eigenvalues converges to a semicircular distribution (Fig.~\ref{Spectral_density_ER} (a))
\begin{equation}
\rho(\lambda) = \begin{cases} (2 \pi \sigma^{2})^{-1} \sqrt{4 \sigma^{2} - \lambda^{2}}~~\mbox{if } |\lambda| < 2\sigma \\
0 ~~ \mbox{otherwise} \end{cases}
\end{equation} 
where $\langle A_{ij}^{2} \rangle$ = $\sigma^{2}$.
Interestingly, this
result matched with a result in random matrix theory about the spectral density of a random matrix, whose elements are Gaussian distributed random numbers, following Wigner's semicircle law \cite{Wigner,Mehta_book}. This finding triggered extensive investigations on different model and real-world networks under the framework of the sophisticated mathematical tool, RMT. RMT was initially proposed by Wigner to explain statistical properties of nuclear spectra and had successful
predictions for the properties of the bulk of the eigenvalues of different complex
systems such as disordered systems, quantum chaotic systems, large complex atoms, etc., followed by numerical and experimental verifications in the last few decades, for instance, analysis of time-series data of stock-market, atmosphere, human EEG, and many more \cite{Mehta_book,Guhr_1998}. The width of the bulk part of the spectrum of ER random networks, i.e. the set of the eigenvalues {$\lambda_{2}, ....... \lambda_{N}$ scales as $\sigma \sqrt{N}$. The edges of the semicircular distribution are known to decay exponentially \cite{Mehta_book}. 

It was found that unlike ER random networks, spectral density of other model and real-world networks do not exhibit semicircular distribution. For the scale-free networks, the spectral density
does not fit the semicircular equation derived by
Wigner, appropriate to the ER random networks. Rather the spectral density of the scale-free networks exhibits a triangle-like structure \cite{Aguiar_2005}. The edges of the triangular distribution of the scale-free networks exhibit a power-law decay. Since the power-law decay is much slower than the
exponential one, the spectrum shows long tails at both edges (Fig.~\ref{Spectral_density_ER} (b)).
\begin{figure}
  \centerline{\includegraphics[width=0.75\columnwidth]{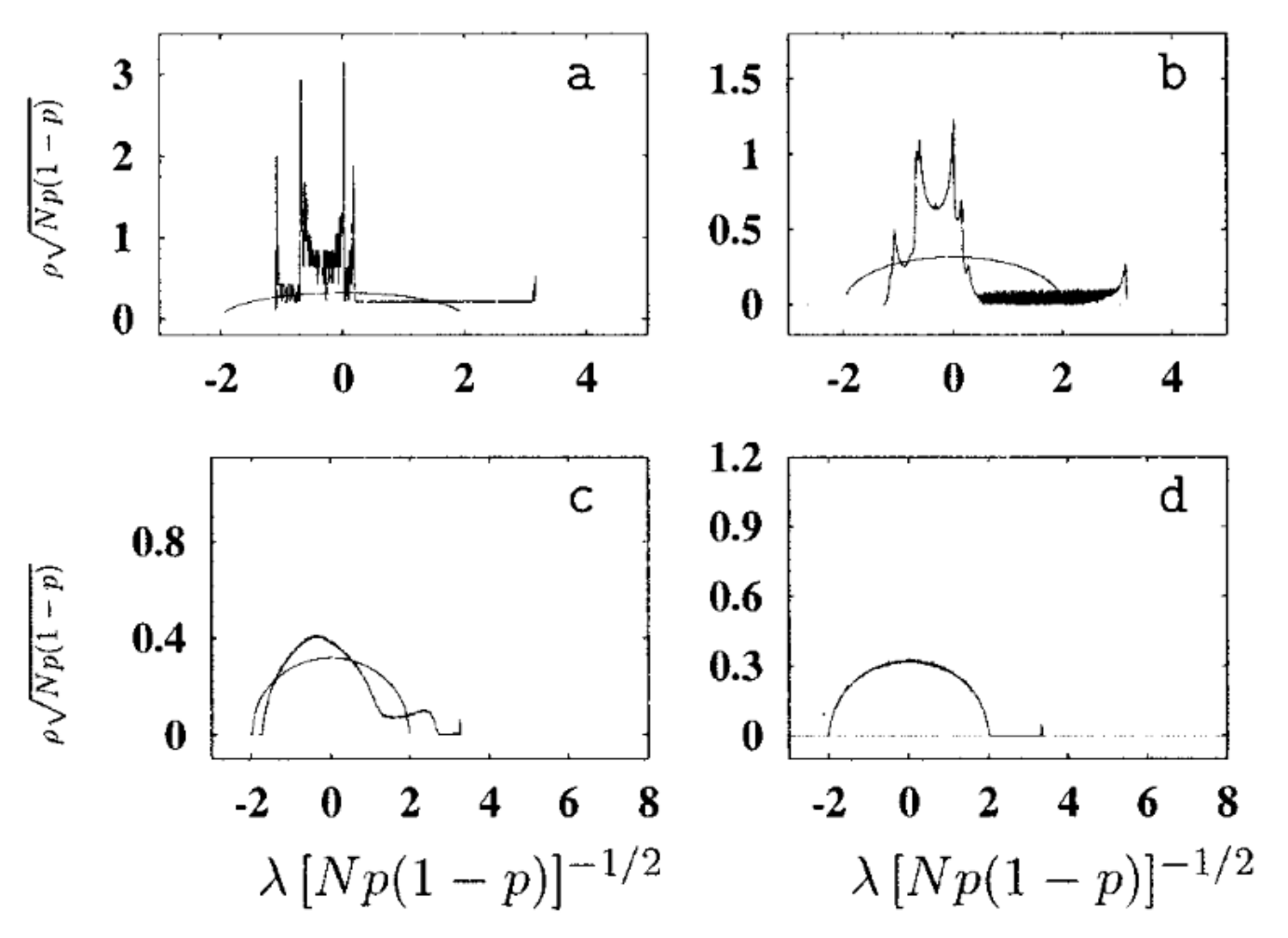}}
  \caption{Spectral densities of small-world networks. Spectral densities of networks at different rewiring probabilities ($p_r$) starting from (a) regular lattice ($p_{r}) = 0$, (b) $p_{r} = 0.01$ (close to small-world transition), (c) $p_{r} = 0.3$ and (d) $p_{r} = 1$ (a random network). The spectral density for a regular lattice as well as for higher rewiring probabilities exhibit sharp peaks, which converges to semicircular structure at $p_{r}$ = 1. The solid line shows the semicircular distribution for comparison. In the plots, 1000 different networks with $N = 1000$ and $\langle k \rangle = 10$ are used for averaging. Reprinted figure (Figure 3) with permission from
I. J. Farkas, I. Der\'enyi, A.-L. Barab\'asi, and T. Vicsek, Phys. Rev. E, 64 (2), 026704 (2001). Copyright (2001) by the
American Physical Society. DOI: \href{https://journals.aps.org/pre/abstract/10.1103/PhysRevE.64.026704}{10.1103/PhysRevE.64.026704}.}
  \label{Spectral_density_SW}
\end{figure}
The tail of the density of eigenvalues ($\rho(\lambda)$) at large $|\lambda|$ is related to the  power-law exponent of the underlying degree distribution ($\gamma$) of the scale-free networks as \cite{Dorogovtsev_PRE_2003,Rodgers_JPA_2005}
\begin{equation}
\rho(\lambda) \sim |\lambda|^{1 - 2\gamma}
\end{equation} 
This expression was shown to be universal for range of scale-free networks, from tree-like networks to dense graphs.
%The density of eigenvalues $\rho(\lambda)$ in the middle part of the spectrum is likely to fit the formula $\rho(\lambda) ~ exp(-|\lambda|/a)$ where $a \approx 1.25$, while the density further out follows the power law $\rho(\lambda) ~ |\lambda|^{-4}$ \cite{Goh_PRE_2001}. 

Another model proposed by Watts and Strogatz to capture the clustering in real graphs was that of the small-world networks. Small-world networks, characterized by high clustering coefficient and low characteristic path length, are generated by rewiring the edges of the 1-d lattice, randomly with a certain rewiring probability ($p_r$). The spectral density of 1-d lattices exhibit sharp peaks (Fig.~\ref{Spectral_density_SW} (a)). At rewiring probabilities close to the small world transition, the {\it blurred} remnants of these peaks persist ((Fig.~\ref{Spectral_density_SW}) (b)-(c)) \cite{Farkas_2001}. However, with further increase in rewiring probabilities, the spectral density plot smoothens and finally converges to a semicircular distribution at $p_r = 1$ (Fig.~\ref{Spectral_density_SW} (d)). It is worth noting here that both Fig.~\ref{Spectral_density_ER} (a) and Fig.~\ref{Spectral_density_SW} (d) depict spectral density of ER random networks. However, in Fig.~\ref{Spectral_density_ER} (a), the spectral density exhibits a peak over the semicircle. The networks in Fig.~\ref{Spectral_density_ER} (a) being sparser than those in Fig.~\ref{Spectral_density_SW} (d), it can be realised that this behaviour arises due to the sparseness of the networks. It has been found that for sparse random networks, in which the number of links grows as the number of nodes, the spectral density does not converge to the semicircle law \cite{Farkas_2001} and exhibits a central peak over the semicircle structure \cite{Kuhn,Farkas_2001}. This central peak pertaining to degenerate eigenvalues is known to provide information about symmetry in underlying networks and hence we have devoted a separate section for discussing degenerate eigenvalues.
\begin{figure}
\centerline{\includegraphics[width=0.75\columnwidth]{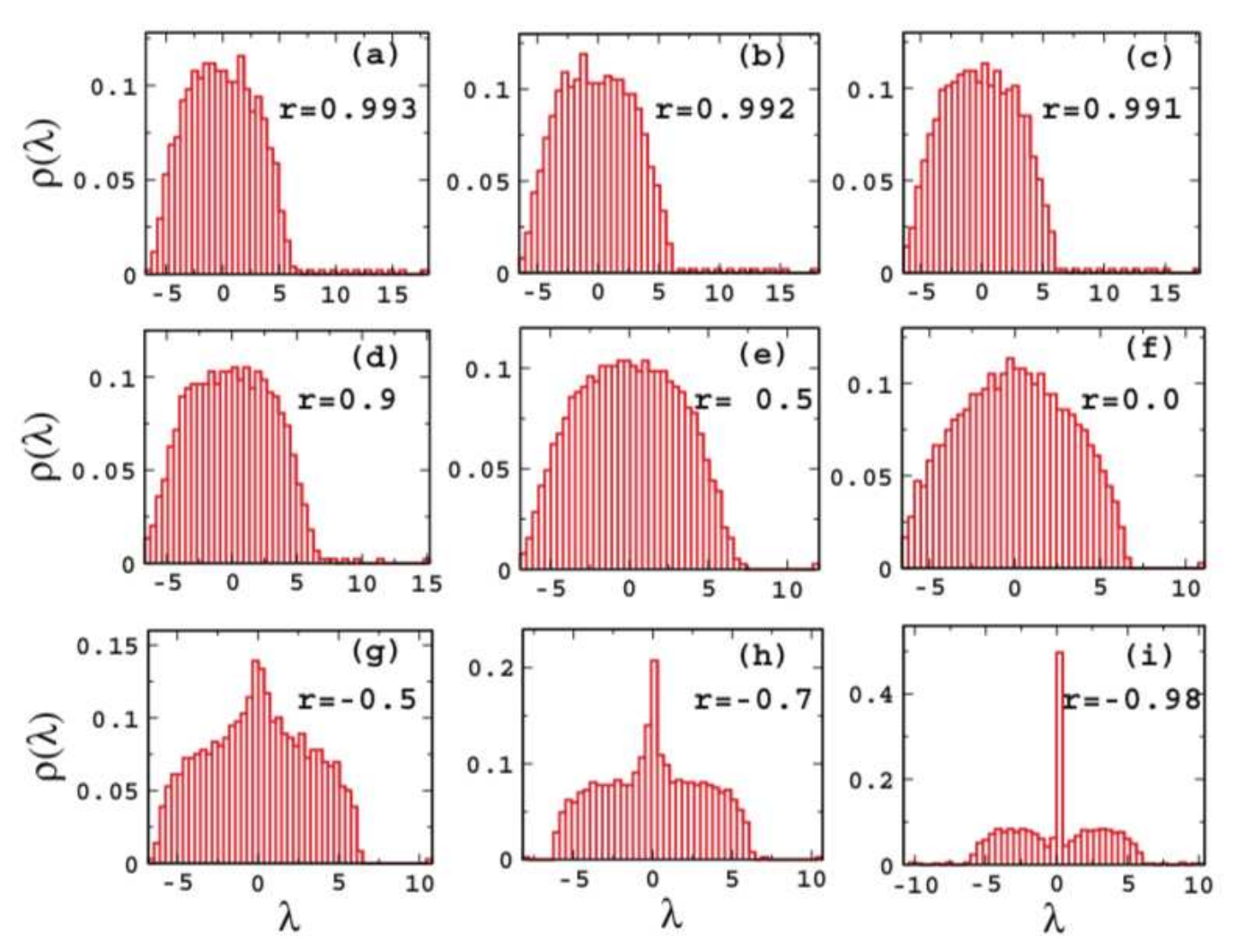}}
\caption{Spectral density for ER random networks with different values of degree-degree correlation coefficient $r$. All graphs are plotted for the networks with size $N = 1000$ and connection probability $p = 0.01$, averaged over 20 different realizations of the networks. Reprinted figure (Figure 1) with permission from
S. Jalan and A. Yadav, Phys. Rev. E 91, 012813 (2015). Copyright (2015) by the
American Physical Society. DOI: \href{https://journals.aps.org/pre/abstract/10.1103/PhysRevE.91.012813}{10.1103/PhysRevE.91.012813}.}
\label{r_lambda}
\end{figure}

A crucial structural property that affects the nature of spectral density is the degree-degree correlations. In case of uncorrelated ER random networks, with an increase in the assortativity, the semicircular distribution remains unchanged (Figs.~\ref{r_lambda} (a)-(e)). As the network is rewired, entailing disassortativity, the spectral density acquires a very different structure from those of the assortative networks. With the overall spectra resembling a double-humped structure, the networks start exhibiting a high degeneracy at zero (Fig.~\ref{r_lambda} (h)), which becomes more pronounced (Fig.~\ref{r_lambda} (i)) as the value of $r$ becomes more negative \cite{SJ_Alok_PRE_2015}. It has been argued out that the presence of bipartite-like or tree-like structures at high disassortativity values of networks might be the reasons behind their high degeneracy at zero \cite{Sokolov,SJ_Alok_PRE_2015}.

Much less was known about the spectra of real-world networks, drawn from complex systems such as the Internet, metabolic pathways, networks of power stations, scientific collaborations, or movie actors, which were inherently correlated and usually very sparse. Increasing availability of data pertaining to such large real-world networks led to investigations on spectral densities of various model and real-world networks \cite{Farkas_2001,Dorogovtsev_PRE_2003,Aguiar_2005,Goh_PRE_2001}. It was found that spectra of such networks were governed by characteristic features of the underlying networks. The spectral density of an array of real-world networks exhibit triangular structure, owing to their underlying scale-free topology \cite{SJ_PhysicaA_2014,SJ_IEEE_2016,SJ_SciRep_2014}. 

Thus, we have seen that the bulk part of the spectrum for model and real networks other than ER random network deviate from RMT prediction in terms of spectral density. In the following, it is demonstrated that inspite of spectral density deviating from RMT prediction, spacing distribution of most of the model and real-world networks follow universal prediction of RMT.

\subsection{Spacing distribution}
For matrices having Gaussian distributed random numbers yielding correlated eigenvalues, the distribution of the spacings between the adjacent eigenvalues, termed as nearest neighbor spacing distribution (NNSD) follows the Wigner-Dyson formula
of Gaussian orthogonal ensemble (GOE) statistics of RMT. On the other hand, for uncorrelated eigenvalues, NNSD
follows Poisson statistics of RMT, which is a property shown by random matrices having nonzero elements only along its
diagonals. The fluctuations in the eigenvalues of adjacency matrices, quantified in terms of NNSD depict crucial properties of the systems \cite{SJ_PRE2007a}. 
\begin{figure}
  \centerline{\includegraphics[width=0.75\columnwidth]{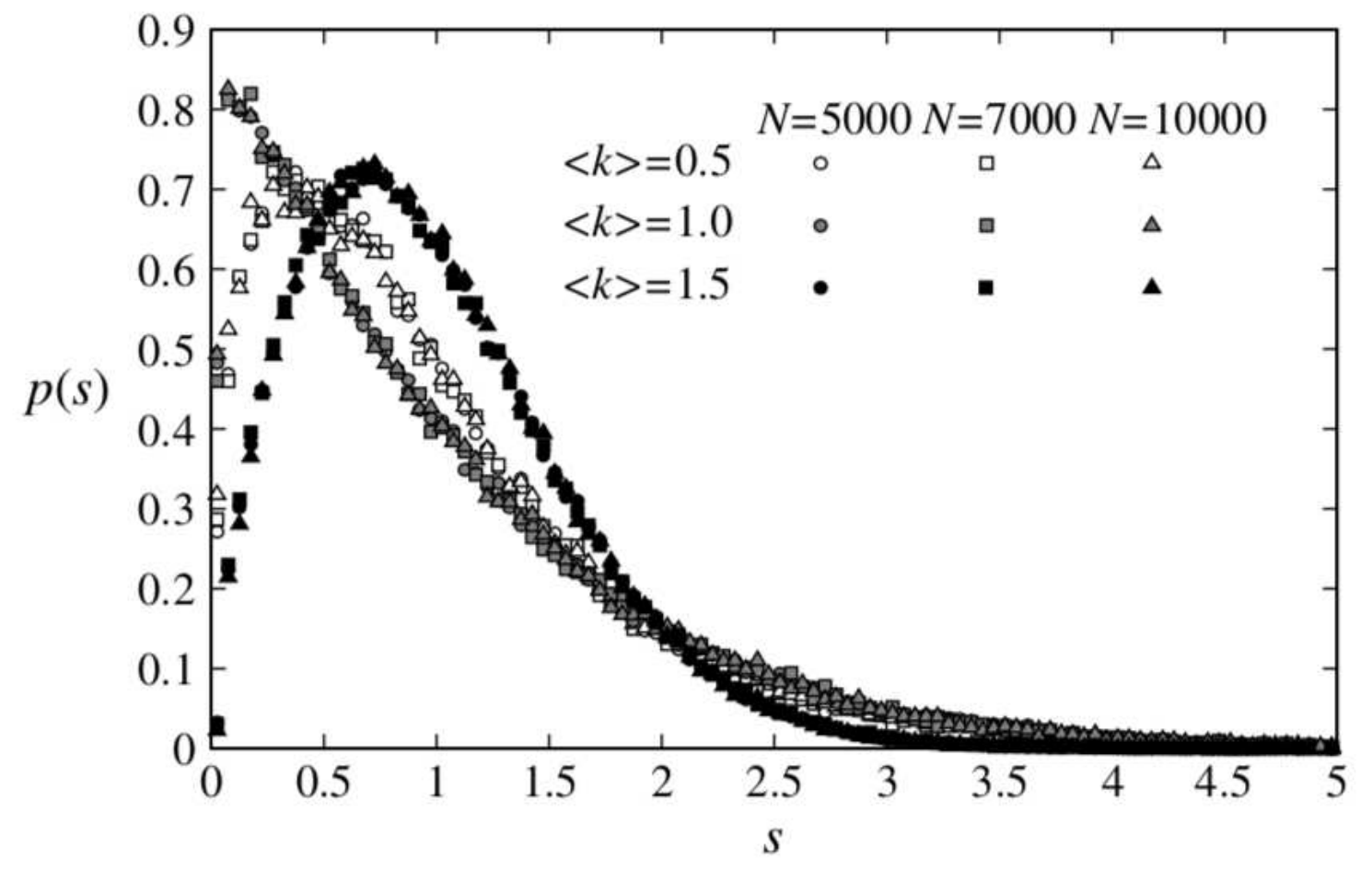}}
  \caption{The nearest neighbour spacing distribution (P(s)) of ER random networks of size $N = 5000$ (circles), $N = 7000$ (squares) and $N = 10000$ (triangles) at average degree $\langle k \rangle = 0.5$ (white symbols), $\langle k \rangle = 1$ (grey symbols) and $\langle k \rangle = 1.5$ (black symbols). The distributions drawn on networks having same average degree coincide with each other. $\textcopyright$ Deutsche Physikalische Gesellschaft. Reproduced figure (Figure 1) by permission of IOP Publishing. CC BY-NC-SA. G. Palla and G. Vattay, Spectral transitions in networks, New J. Phys. 8 (12), 307 (2006). DOI: \href{http://iopscience.iop.org/article/10.1088/1367-2630/8/12/307/meta}{10.1088/1367-2630/8/12/307}.}
  \label{NNSD_k}
\end{figure}

For calculating nearest neighbour spacings, it is customary in RMT to unfold the eigenvalues by a transformation 
$\overline{\lambda_{i}} = \overline{{N}}(\lambda_{i})$ in order to get universal properties of the fluctuations 
of eigenvalues, where $\overline{N}$ is average integrated eigenvalue density \cite{Mehta_book}. Since there does not exist any analytical form for $\overline{N}$, the spectrum is numerically unfolded by polynomial curve fitting \cite{Mehta_book}. After unfolding, average spacing 
becomes unity, independent of the system. Using the unfolded
spectra, spacings are calculated as $s_1(i) = {\overline{\lambda}}_{i+1}-{\overline{\lambda}}_{i}$.
In the case of GOE statistics, the nearest neighbor spacing distribution is denoted by  
\begin{equation}
P(s) = \frac{\pi}{2}s\exp\left(-\frac{{\pi}s^2}{4}\right)
\label{eq_NNSD}
\end{equation}
\begin{figure} 
  \centerline{\includegraphics[width=0.7\columnwidth]{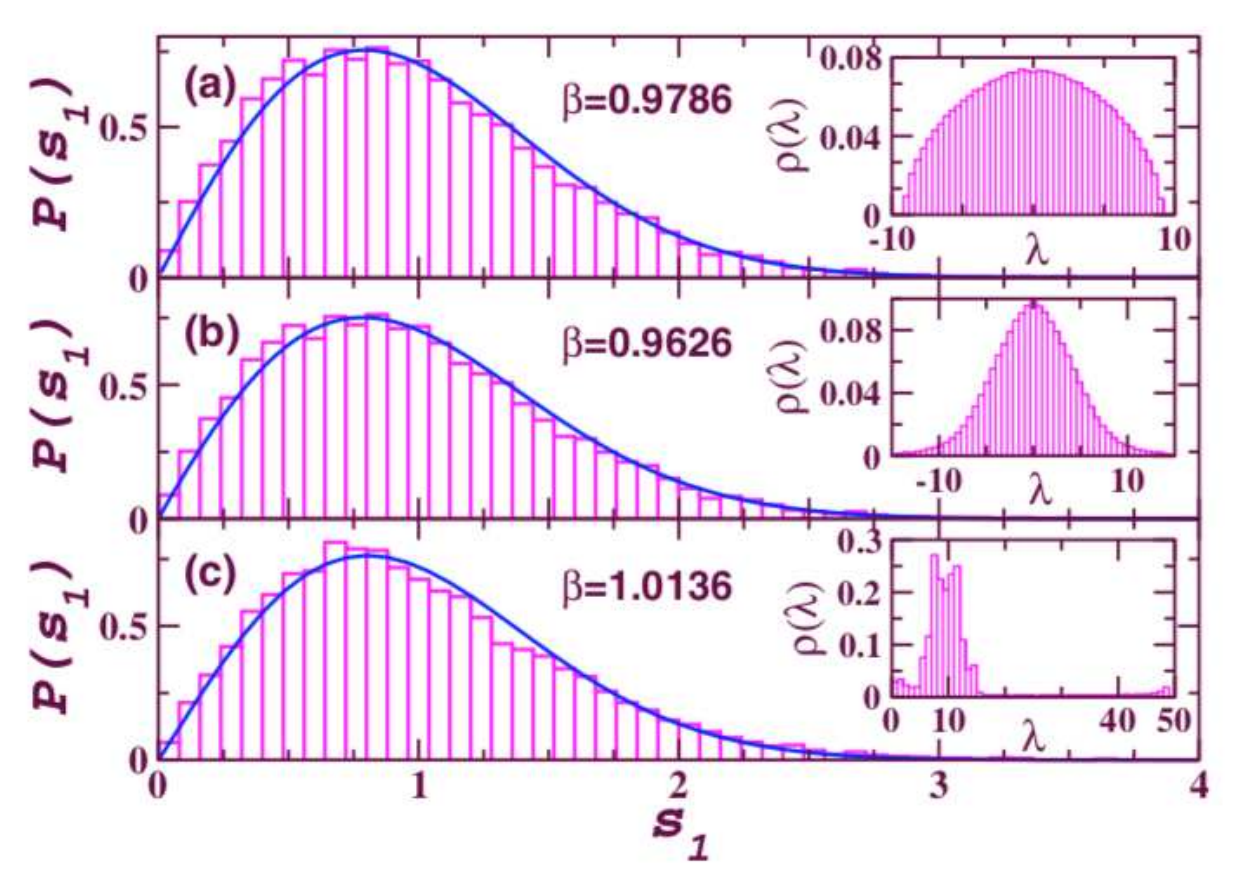}}
  \caption{Nearest neighbour spacing distribution of the adjacency matrices of (a) ER random network, (b) scale-free network and (c) small-world network. The histograms are numerical results and the solid lines represent the fitted Brody distribution (Eq.~\ref{Brody_eq}). Figures are plotted for the average over 10 random realizations of the networks having size $N=2000$ and average degree $\langle k \rangle =20$. Insets show respective spectral densities ($\rho(\lambda)$). Note that $P(s)$ and $P(s_1)$ are interchangeably used to represent NNSD as per the choice of the authors. Reprinted figure (Figure 1) with permission from
S. Jalan and J. N. Bandyopadhyay, Phys. Rev. E 76 (4), 046107 (2007). Copyright (2007) by the
American Physical Society. DOI: \href{https://journals.aps.org/pre/abstract/10.1103/PhysRevE.76.046107}{10.1103/PhysRevE.76.046107}.}
  \label{NNSD_SpectralDensity_SJ_PRE}
\end{figure}

For intermediate cases, the spacing distribution is described by Brody parameter \cite{Brody}.
\begin{subequations}
\begin{align}
P_{\beta}(s)&=as^\beta\exp\left(-\alpha s^{\beta+1}\right)
\label{Brody_eq}
\end{align}
where $a$ and $\alpha$ are determined by the parameter $\beta$ as follows:
\begin{align}
A&=(1+\beta)\alpha, \, \, \, \alpha=\left[{\Gamma{\left(\frac{\beta+2}{\beta+1} \right)  }}\right]^{\beta+1}
\end{align}
\label{eq_brody}
\end{subequations}
\begin{figure} 
  \centerline{\includegraphics[width=0.49\columnwidth]{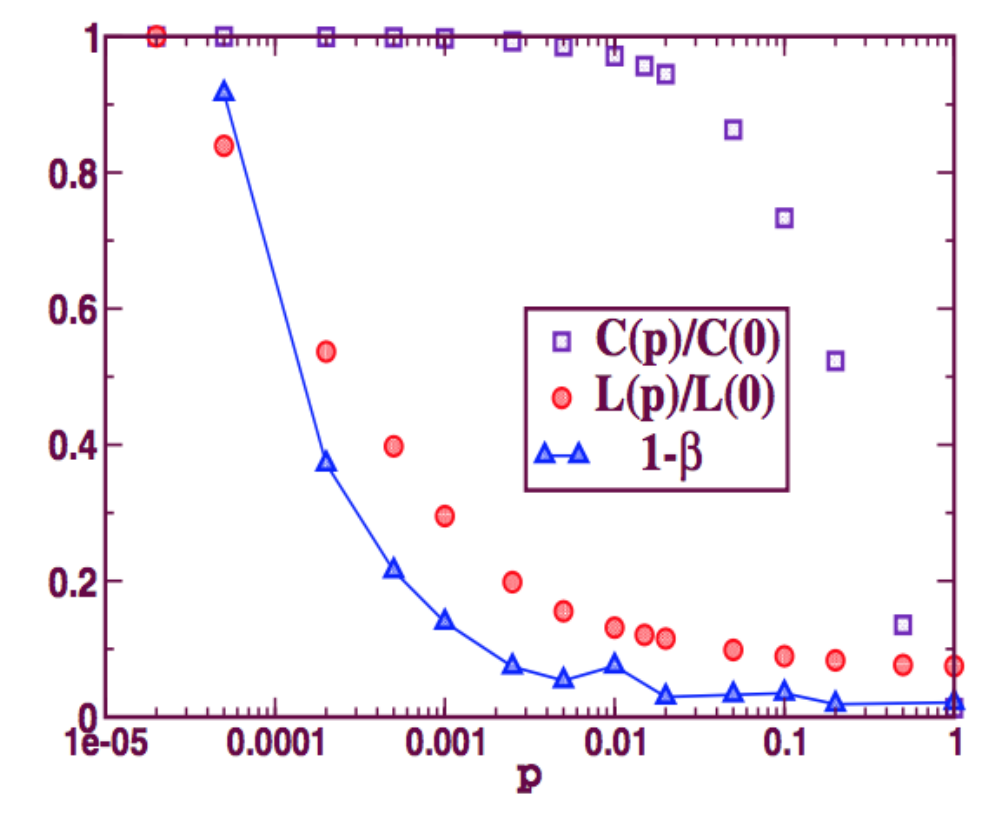}}
  \caption{The shifted Brody parameter $1-\beta$ ($\triangle$) plotted as a function of normalized characteristic path length ($\circ$) and normalised clustering coefficient ($\Box$). Reprinted figure (Figure 4) with permission from
J. N. Bandyopadhyay and S. Jalan, Phys. Rev. E, 76 (2), 026109 (2007). Copyright (2007) by the
American Physical Society. DOI: \href{https://journals.aps.org/pre/abstract/10.1103/PhysRevE.76.026109}{10.1103/PhysRevE.76.026109}.}
  \label{Beta_PathLength}
\end{figure}
This is a semi-empirical formula characterized by parameter $\beta$. As $\beta$ goes from 0 to 1, the
Brody distribution smoothly changes from Poisson to GOE. The spacing distributions of different
networks are fitted by the Brody distribution $P_{\beta}(s)$. This fitting gives an estimation of $\beta$, and
consequently identifies whether the spacing distribution of a given network is Poisson, GOE, or the
intermediate of these two \cite{Brody}.
The spectral density of ER random networks and of Gaussian distributed random matrices were both found to be semicircular, so it was expected that their spacing distributions would be identical. Fitting the NNSD (Eq.~\ref{eq_NNSD}) of ER random network with the Brody distribution (Eq.~\ref{eq_brody}) yielded the value of Brody parameter close to 1. This value of $\beta \sim 1$ clearly indicates that as expected, the NNSD of ER random networks follows GOE statistics of RMT. 
\begin{figure}
\centerline{\includegraphics[width=0.7\columnwidth]{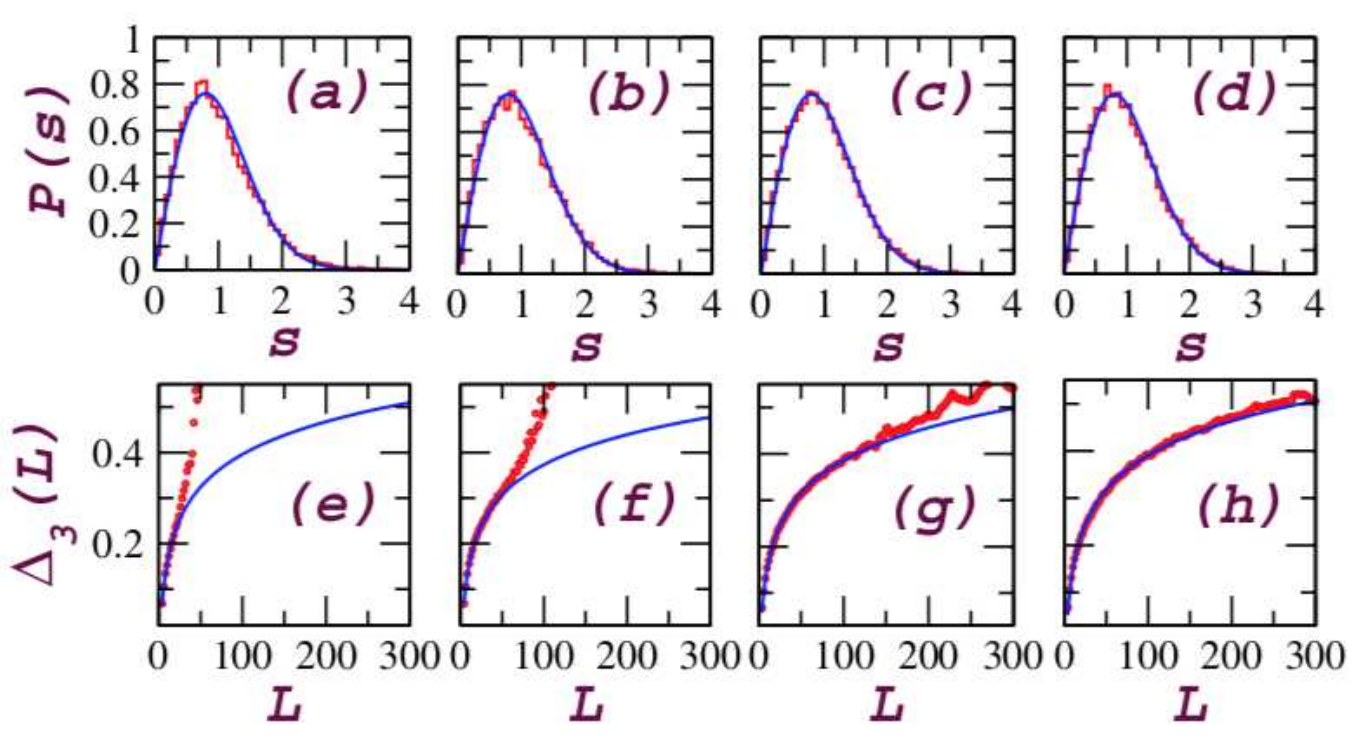}}
\caption{Change in the spectral behavior for different values of rewiring probability, $p_r$. All these plots are drawn on networks having size $N = 2000$ and average degree $\langle k \rangle = 20$. (a)-(d) present plots of NNSD for different values of $p_r$, where histograms depict NNSD of network spectra and the solid curves depict the theoretical prediction of NNSD for GOE. (e)-(h) plots $\Delta_3$ statistic for different values of $p_r$, where the open circles depict $\Delta_3$ statistic of network spectra and the solid curves depict the theoretical prediction of $\Delta_3$ statistic for GOE. $p_r = 0.002$ for plots (a) and (e); $p_r = 0.02$ for plots (b) and (f); $p_r = 0.05$ for plots (c) and (g) and $p_r = 0.1$ for plots (d) and (h). It was found that for different values of $p_r$, the NNSD follows GOE statistics ((a)-(d)), indicating universality in short range correlations. However, plots of $\Delta_3$ statistic demonstrate that the length $L_0$ upto which the spectra follows GOE statistics increases with increase in the value of $p_r$, i.e. with increase in randomness. Reproduced Figure 2 with permission from S. Jalan and J. N. Bandyopadhyay, Randomness of random networks: A random matrix analysis, EPL 87, 48010 (2009). DOI: \href{http://iopscience.iop.org/article/10.1209/0295-5075/87/48010/pdf}{10.1209/0295-5075/87/48010}.}
\label{NNSD_Delta3_SJ_EPL}
\end{figure}

The shape of the distribution of nearest neighbour spacings has been shown to be determined by the average degree of the underlying network \cite{Vattay_NJP_2006}. The NNSD of ER random networks of different sizes but having the same average degree, exhibit the same shape of the probability distribution curve (Fig.~\ref{NNSD_k}). Around the critical point of the percolation transition in case of ER random networks at $\langle k \rangle = 1$, the NNSD undergoes a significant change when $\langle k \rangle$ is varied. At $\langle k \rangle = 1$, the NNSD follows Poisson distribution, while for $\langle k \rangle \gg 1$, the NNSD abides by GOE statistics \cite{Vattay_NJP_2006}.

There are inherent differences of other model networks from ER random networks in
terms of various local and global properties, for instance as depicted through their spectral densities (insets of Fig.~\ref{NNSD_SpectralDensity_SJ_PRE}). But interestingly, the fluctuations of the eigenvalues of adjacency matrices quantified in terms of nearest neighbor spacing distribution of scale-free and small-world networks also followed GOE statistics \cite{SJ_PRE2007a}. More interestingly, an analogy was drawn between the onset of small-world transition, quantified by the structural properties of
networks namely the diameter and the clustering coefficient, and the transition from Poisson to GOE statistics, quantified by Brody
parameter \cite{Brody} characterizing a spectral property. It was found that the NNSD
changes with the transition of a regular lattice to a small-world network. According to the Watts-Strogatz model, the small-world networks are constructed by rewiring the regular network with a probability (say, $p_r$), which generates a network with $N \langle k \rangle p_r$ random connections, without altering the number of nodes or edges. It is found that for the regular network ($p_r = 0$), NNSD follows Poisson statistics, while for $p_r = 1$ i.e. when all
the edges of the regular lattice are randomly rewired,
NNSD follows the GOE statistics.
\begin{figure} 
  \centerline{\includegraphics[width=0.5\columnwidth]{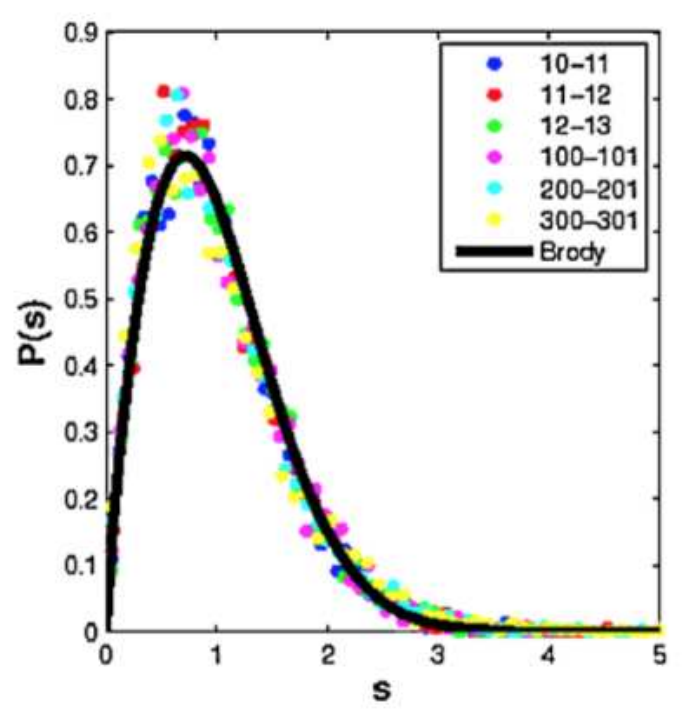}}
  \caption{Nearest neighbor spacing distribution ($P(s)$) of the displacement covariance matrix ensembles drawn on protein energy levels. The displacement covariance matrix is obtained from molecular dynamics simulations carried out on {\it E. Coli} adenylate kinase, a $N = 214$ amino acid single-chain phosphotransferase protein. The distribution of spacings emulate GOE statistics, the circles being the numerical results and the solid line representing fitted Brody distribution (Eq.~\ref{eq_brody}). The value of $\beta$ close to $1$ corresponds to the GOE distribution. Reprinted figure (Figure 4 right panel) with permission from
R. Potestio, F. Caccioli, and P. Vivo, Phys. Rev. Lett. 103, 268101 (2009). Copyright (2009) by the American Physical Society. DOI: \href{https://journals.aps.org/prl/abstract/10.1103/PhysRevLett.103.268101}{10.1103/PhysRevLett.103.268101}.}
  \label{NNSD_Amino}
\end{figure}
At a very small rewiring probability ($p_r$), the transition to the 
small-world network happens, which is marked by a drastic decrease in the characteristic path length while still having a high clustering coefficient. Interestingly, it was shown that the NNSD changes from Poisson to GOE with a very small increment in $p_r$, and the transition to GOE takes place exactly at the onset of small-world transition \cite{SJ_PRE2007a}. As depicted in Fig.~\ref{Beta_PathLength}, the shifted $\beta$ (1 - $\beta$) and the normalized characteristic length were found to display similar trends and strong correspondence, thus establishing the significance of the Brody parameter (a spectral attribute) in capturing {\it randomness} in networks (a structural feature). The NNSD, when plotted for increasing values of $p_r$, starting from the small-world transition point, followed universal GOE statistics for all values of $p_r$ (Fig.~\ref{NNSD_Delta3_SJ_EPL} (a)-(d)) \cite{SJ_EPL_2009}. In case of random geometric networks as well, high values of connection radius yielded the NNSD of the underlying networks to follow the universal GOE statistics of RMT \cite{Knight_EPL_2017}.

Unlike spectral density which turned out to be a distinguishing feature between random and real-world networks, the nearest neighbour spacing distribution of most of the real-world networks were found to follow the universal GOE statistics of RMT.  
Many of the different biological networks, such as protein-protein interaction (PPI) network in budding
yeast \cite{SJ_PRE2007a}, {\it C. elegans}, {\it D. melanogaster},
 {\it H. pylori}, {\it H. sapiens}, {\it S. cerevisiae} and {\it E. coli} \cite{SJ_PhysicaA_2014}, breast cancer PPI networks \cite{SJ_SciRep_2014}, gene co-expression networks
of Zebrafish \cite{SJ_EPL_2012}, Alzheimer's disease \cite{SJ_PRE_2010}, yeast \cite{Luo_2007} all fall under the universality class of Gaussian orthogonal ensemble (GOE) statistics. The NNSD of covariance matrices of amino acid displacements also follow universal GOE statistics of RMT \cite{Potestio_PRL_2009}, depicted in Fig.~\ref{NNSD_Amino}. Social networks, such as the movie co-actor networks \cite{SJ_Plosone_2014}, the networks of scientific collaborators \cite{SJ_IEEE_2016} have been found to follow RMT prediction for distribution of nearest neighbour spacings. The NNSD of the networks following GOE statistics indicates {\it minimal amount of 
randomness} in the underlying networks bringing upon correlations between only nearest
neighbours in spectra. 
The random geometric networks have been show to display a parameter-dependent transition between the GOE statistics for high values of connection radius and closer to Poisson statistics for low values of connection radius \cite{Knight_EPL_2017}.
The NNSD has also been used to determine the correlation
threshold for identifying gene co-expression networks \cite{Luo_2007}.
However, the distribution of nearest neighbour spacings tells about only the short range correlations in eigenvalues \cite{SJ_PRE2007b}. The NNSD captures the universality present in a wide range of complex networks. However, $\beta$ being close to $1$ for all networks having $p_r$ greater than that of small-world transition, as well as for scale-free networks and real-world networks, the NNSD fails to capture differences in these networks. The spectral rigidity of the networks calculated in terms of $\Delta_3 (L)$ statistic probes for long range correlations in eigenvalues, and unravels differences in complex networks. 

\subsection{$\Delta_3$ statistic}
\begin{figure} 
  \centerline{\includegraphics[width=0.8\columnwidth]{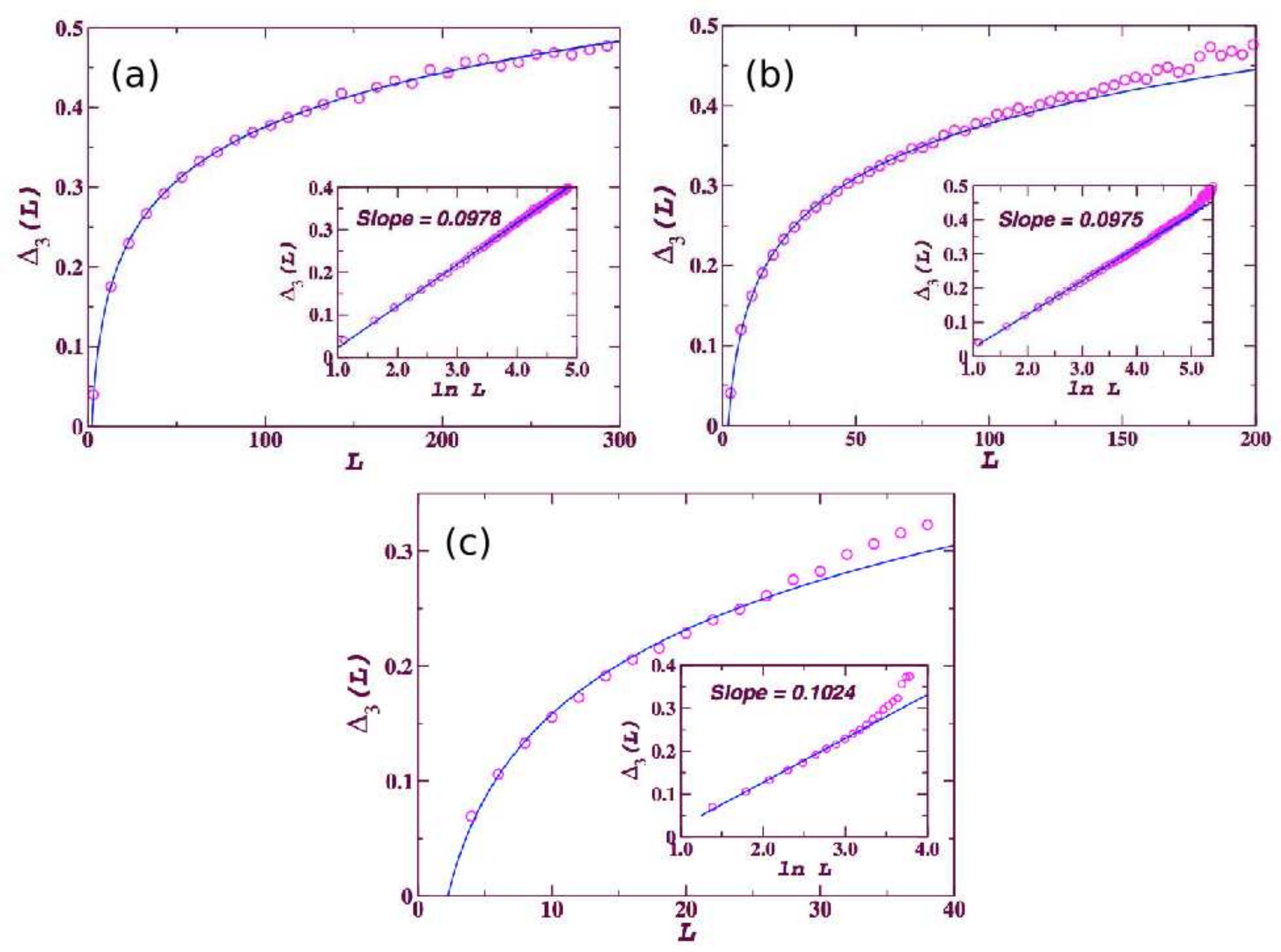}}
  \caption{$\Delta_3$(L) statistics (open circles) for
(a) ER random network, (b) scale-free network and (c) small-world network of size $N = 2000$ and $\langle k \rangle = 20$, ensemble over ten networks. The solid line represents the GOE prediction.  
$\Delta_3 (L)$ statistic for the three different model networks abide by the RMT prediction for varying lengths ($L_0$). Reprinted figures (Figures 3, 4 and 5) with permission from S. Jalan and J. N. Bandyopadhyay, Phys. Rev. E 76 (4), 046107 (2007). Copyright (2007) by the American Physical Society. DOI: \href{https://journals.aps.org/pre/abstract/10.1103/PhysRevE.76.046107}{10.1103/PhysRevE.76.046107}.}
  \label{Delta3_model}
\end{figure}
The $\Delta_3 (L)$ statistic measures the least-square deviation of the spectral staircase function representing
the cumulative density $N(\overline{\lambda})$ from the best fitted straight line for
a finite interval of length $L$ of the spectrum given by
\begin{equation}
\Delta_3 (L;x) = \frac{1}{L} min_{a,b}   \int_x^{x+L} [N(\overline{\lambda})-a\overline{\lambda}-b]^2 d\overline{\lambda}
\label{eq_delta3}
\end{equation}
where $a$ and $b$ are regression coefficients obtained after least square fit. Average over several choices
 of x gives the spectral rigidity $\Delta_3 (L)$. For GOE case, $\Delta_3 (L)$ depends logarithmically on L, i.e.
\begin{equation}
\Delta_3(L)  \sim \frac{1}{\pi^2} \ln L. 
\label{eq_delta3_goe}
\end{equation}

The $\Delta_3(L)$ statistic of the spectrum of the network provides information
of long range correlations in the eigenvalues.
For the ER random networks, the spectral rigidity, quantified in terms of $\Delta_3(L)$ statistic (calculated as per Eq.~\ref{eq_delta3}) agrees very well with the RMT predictions (given by Eq.~\ref{eq_delta3_goe}) up to a very large extent (Fig.~\ref{Delta3_model} (a)). The expected linear behaviour of $\Delta_3(L)$ with a slope very close to the RMT predicted value,
i.e. $\frac{1}{\pi^2} \sim 0.1013$ (Eq.~\ref{eq_delta3_goe}) is observed. This is not surprising as adjacency matrices of random networks have 1's distributed randomly in the networks and hence statistical properties of unfolded eigenvalues of these matrices behave in similar manner as those of random matrices. The deviation from the universal GOE statistic arises due
to finite size of adjacency matrices of networks. Further, for scale-free networks as well, 
the $\Delta_3(L)$ statistic agrees very well with that of the RMT predictions up to a sufficiently large extent with the slope being very close to $\frac{1}{\pi^2}$ (Fig.~\ref{Delta3_model} (b)). This confirms that scale-free networks follow universal GOE statistics, however,
the extent upto which it follows the GOE statistics is much lesser than that of the ER random networks. This is again not surprising that scale-free networks are less random
than the ER random networks. The scale-invariant power law behaviour of scale-free networks indicating self-organization in the underlying networks can be attributed as a reason behind lesser randomness as compared to ER random networks, captured through the $\Delta_3(L)$ statistic. As expected, the $\Delta_3(L)$ statistic of the small-world network follows RMT prediction for a certain extent, but much less than that of the ER random 
and scale-free networks \cite{SJ_PRE2007b}, presented in Fig.~\ref{Delta3_model} (c). 
The universal GOE statistics followed by the small-world networks suggests that
these networks have some amount of randomness, arising due to random rewiring of the regular
lattices, however, deviation from the universal GOE statistics at very small value
of $L$ implies that besides randomness, small-world networks possess 
system-specific features such as abundance of the clique structure.
Upon a systematic calculation of $\Delta_3(L)$ statistic for networks having different values of rewiring probabilities ($p_r$), one can witness a perfect correlation between the value
of $L$ for which spectra follows GOE statistic and amount of {\it randomness} present in the
network. Starting from the small-world transition point, on increasing the probability of random rewiring, it has been realized that $\Delta_3(L)$ statistic displays a constant increase in the value of $L_0$, the extent upto which the $\Delta_3(L)$ statistic follows RMT prediction (Fig.~\ref{NNSD_Delta3_SJ_EPL} (e)-(h)) \cite{SJ_EPL_2009}. The $\Delta_3(L)$ statistic following RMT prediction for a larger range as the networks approach a random network ($p_r = 1$) is again a demonstration that the $\Delta_3(L)$ statistic captures randomness in networks.

Further, degree-degree correlations of underlying networks has also 
been seen to affect their $\Delta_3(L)$ statistic. As assortativity of a network is 
decreased, the randomness of the network has been demonstrated to show an increases \cite{SJ_Alok_PRE_2015}. This increase in randomness continues until the underlying networks are neutral ($r$=$0$), supporting the idea that the network reaches maximum randomness. The extent to randomness, measured in terms of $L_{0}$ then remains steady for a further
decrease in the value of $r$ to its smallest possible
value, i.e. to the maximum disassortativity  \cite{SJ_Alok_PRE_2015}. These analyses based on different model networks has enabled the network science community to understand randomness and complexity of real-world networks. In the following we review the results drawn from
$\Delta_3$ analysis of real-world networks.

The long range correlations in eigenvalues of different real-world networks, measured by $\Delta_3 (L)$ statistic, have been show to agree well with the RMT predictions up to different length $L_0$ \cite{SJ_PhysicaA_2014,SJ_IEEE_2016}.  The spectra of real-word networks
 analyzed using $\Delta_3$ statistic span range from biological, technological and social systems. For instance, non-consistent changes in the brain networks have also been captured by the long range correlations in eigenvalues \cite{Wang_Chaos_2015}.
Again the universal behaviour of eigenvalues for different real-world networks
suggest existence of certain amount of {\it randomness} in all these systems. On other hand,
deviation from the universality upto a certain value of $L_0$ indicates presence of order in
these systems. A different value of $L_0$ for which the individual network follows
RMT indicates varying amounts of randomness in the underlying networks. Presence
of universal and non-universal behaviour of $\Delta_3$ statistic furnishes the importance of randomness and order in the sustenance of the real-world systems. 
Using this relationship between amount of {\it randomness} in networks and
the value of $L_0$ for which the spectra follow GOE statistic, it was shown
that the normal and diseased states of breast cancer network have varying amounts of randomness among the two states \cite{SJ_SciRep_2014}. The fact that $\Delta_3 (L)$ statistic defies RMT prediction for few of the real-world networks \cite{SJ_PhysicaA_2014,SJ_IEEE_2016}, has been interpreted as the existence of a very {\it minimal amount of randomness} in the underlying matrices bringing upon correlations between only nearest neighbours in spectra. 

\section{Degenerate eigenvalues}
\begin{figure}
  \centerline{\includegraphics[width=0.8\columnwidth]{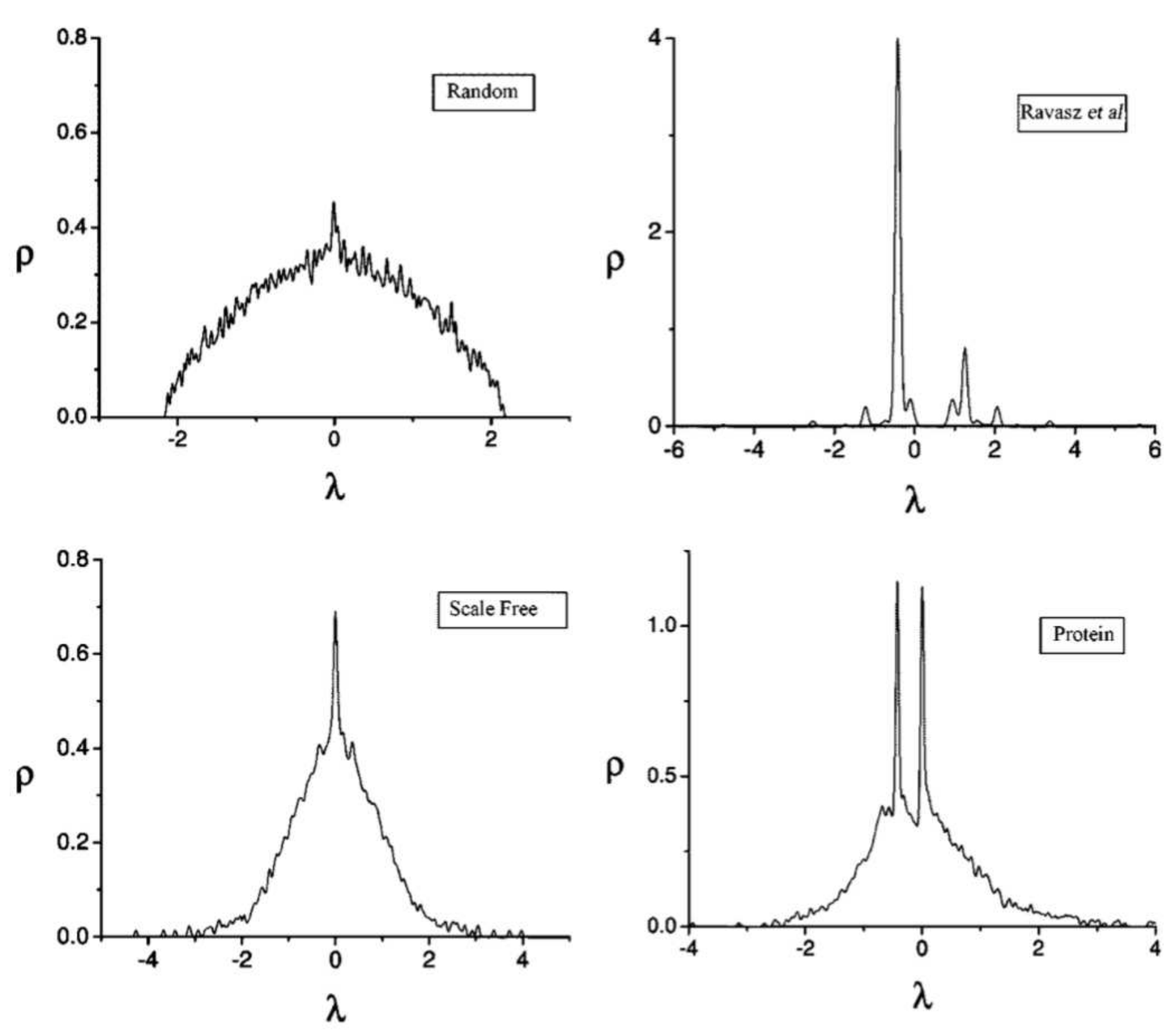}}
  \caption{Plots of spectral density for a random, scale-free, hierarchical network of Ravasz {\it et al.} \cite{Ravasz_Science_2002} (all with network size 1024) and the protein-protein interaction network (with 1297 nodes). Reprinted figure (Figure 1) with permission from
M. A. M. de Aguiar and Y. Bar-Yam, Phys. Rev. E 71, 016106 (2005). Copyright (2005) by the
American Physical Society. DOI: \href{https://journals.aps.org/pre/abstract/10.1103/PhysRevE.71.016106}{10.1103/PhysRevE.71.016106}.}
  \label{Aguiar_Degenerate}
\end{figure}
The spectra of many real world networks exhibit properties which are different from those of random networks generated using various models. One such property is the existence of a very high degeneracy at the zero eigenvalue. Unlike the spectral density of ER random networks which follow the Wigner's semicircle law, the spectral density of scale-free networks exhibit a triangular structure with a peak at zero eigenvalue (Fig.~\ref{Aguiar_Degenerate}).
Another category of networks, the hierarchical networks proposed by Ravasz {\it et al.} to explain the self-similarity in biological systems \cite{Ravasz_Science_2002}, fail to exhibit a high degeneracy at zero eigenvalue. However, almost all biological and technological networks exhibit high degeneracy at the zero eigenvalue \cite{SJ_PhysicaA_2014,Dorogovtsev_PRE_2003,SJ_EPL_2015b,SJ_EPJB_2015} and as depicted in Fig.~\ref{Aguiar_Degenerate}. The gene duplication, a basic mechanism behind the growth of many biological systems \cite{Teichmann}, has been suggested as one of the reasons behind the occurrence of high degeneracy at zero eigenvalue in their underlying networks \cite{Kamp_PRE_2005}. However, from a very simple matrix algebra calculation, it has been shown how duplication of nodes leads to lowering of the rank of the corresponding matrix, hence contributing one additional zero eigenvalue in the spectra \cite{Jost,SJ_Chaos_2015}. 

A theorem \cite{linearBook} relating the degeneracy at zero eigenvalues with the properties of the matrix states that for an adjacency matrix of size $N$ and rank $r$ there will be exactly $N-r$ zero eigenvalues. Therefore, given the rank of an 
adjacency matrix, the degenerate zero eigenvalues can be derived. Factors responsible for lowering of the rank of an adjacency matrix are enlisted in the following:
 
(a) Complete row (column) duplication: When two rows (columns) have exactly same entries,
\begin{equation}
R_1=R_2
\label{complete}
\end{equation}
subtracting one such row from the other yields one of the rows to attain all zero values, thus reducing the rank of the matrix by one.
\begin{figure}
\centerline{\includegraphics[width=0.6\columnwidth]{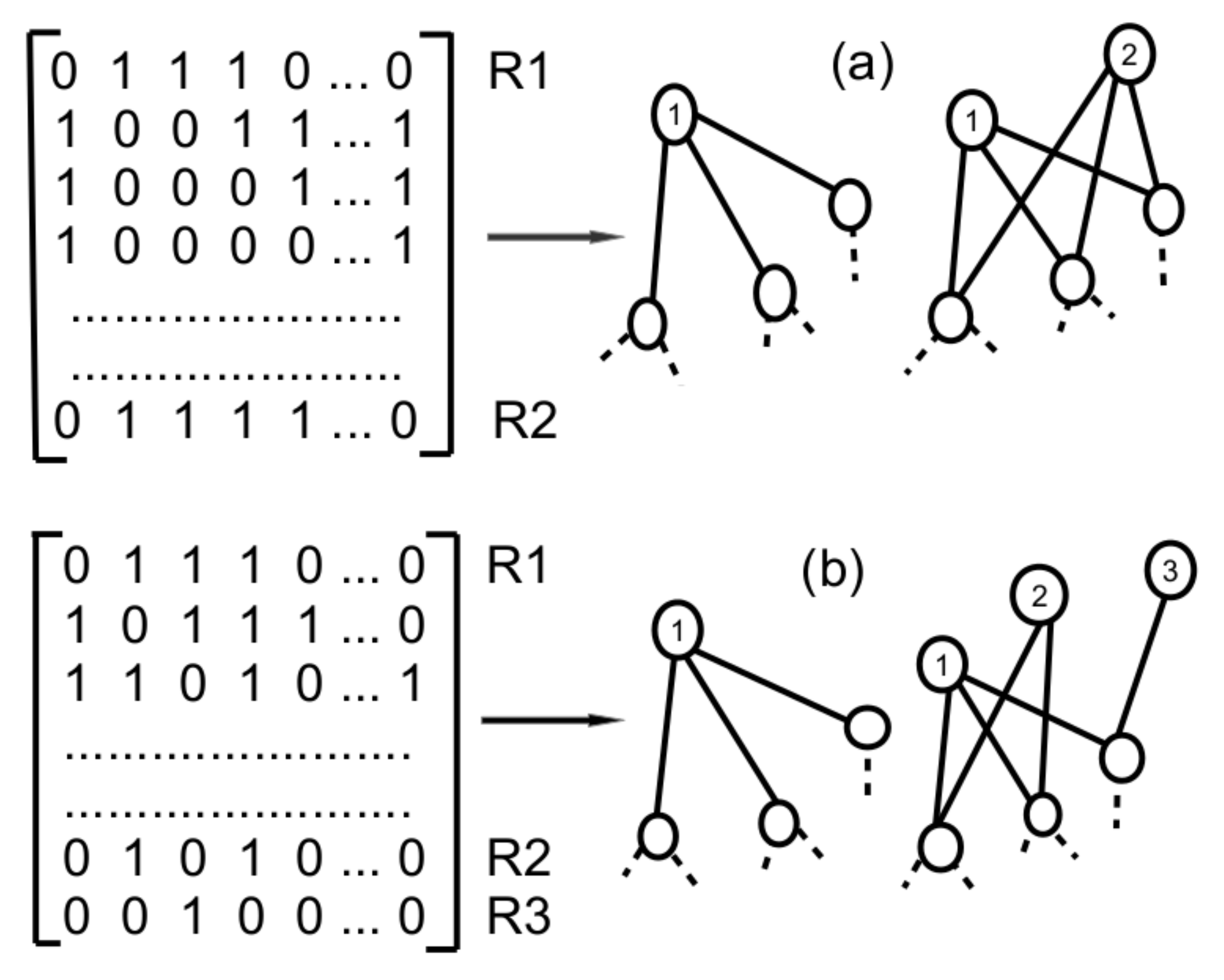}} 
\caption{Schematic diagram
representing (a) complete node duplication (Eq.~\ref{complete}) and (b) partial node duplication (Eq.~\ref{partial}) in networks. Reproduced figure (Figure 1) from A. Yadav and S. Jalan, Chaos 25, 043110 (2015), with the permission of AIP Publishing. DOI: \href{https://aip.scitation.org/doi/10.1063/1.4917286}{10.1063/1.4917286}.}
\label{Schematic_duplication}
\end{figure}

(b) Partial row (column) duplication: When two or more rows (columns) added together have exactly same entries as some other row or column,
\begin{equation} 
R_1=R_2+R_3, \, \, \, {\text or} \, \, \, R_1+R_2=R_3+R_4+R_5
\label{partial}
\end{equation}
it leads to duplication of an existing node.

(c)An isolated node in the network leads to all zero entries in the corresponding row and column, thus lowering the rank of the matrix by one.

Conditions (a) and (b) lead to linear dependence of row (column), reducing the rank of the matrix. There are $N(N-1)/2$ possible ways in which condition (a) of complete duplication can be realized, while for the partial duplication (b) among `$x$' number of nodes with `$y$' number of nodes, there can be $\frac{N!}{2.(N-x-y)!}$ possibilities. Hence, for a given network, checking the existence of condition (b) becomes computationally exhaustive as with increase in network size the number of possibilities becomes very large. 

In order to demonstrate the effect of duplication on zero degeneracy, considering ER random network, (i) a new node was added to the existing network in a way that it satisfies the complete node
duplication criteria (condition (a)); (ii) two new nodes were added to the existing network
in a way that in coalition they mimic the neighbors of an existing node (condition (b)). It was found that with entry of every new node in the network satisfying conditions (a) or (b) of complete or partial duplication, there was an addition of exactly one zero eigenvalue in the spectra \cite{SJ_Chaos_2015}. The number of duplicates (complete or partial) equals the number of zero eigenvalues.
 \begin{figure}
\centerline{\includegraphics[width=0.7\columnwidth]{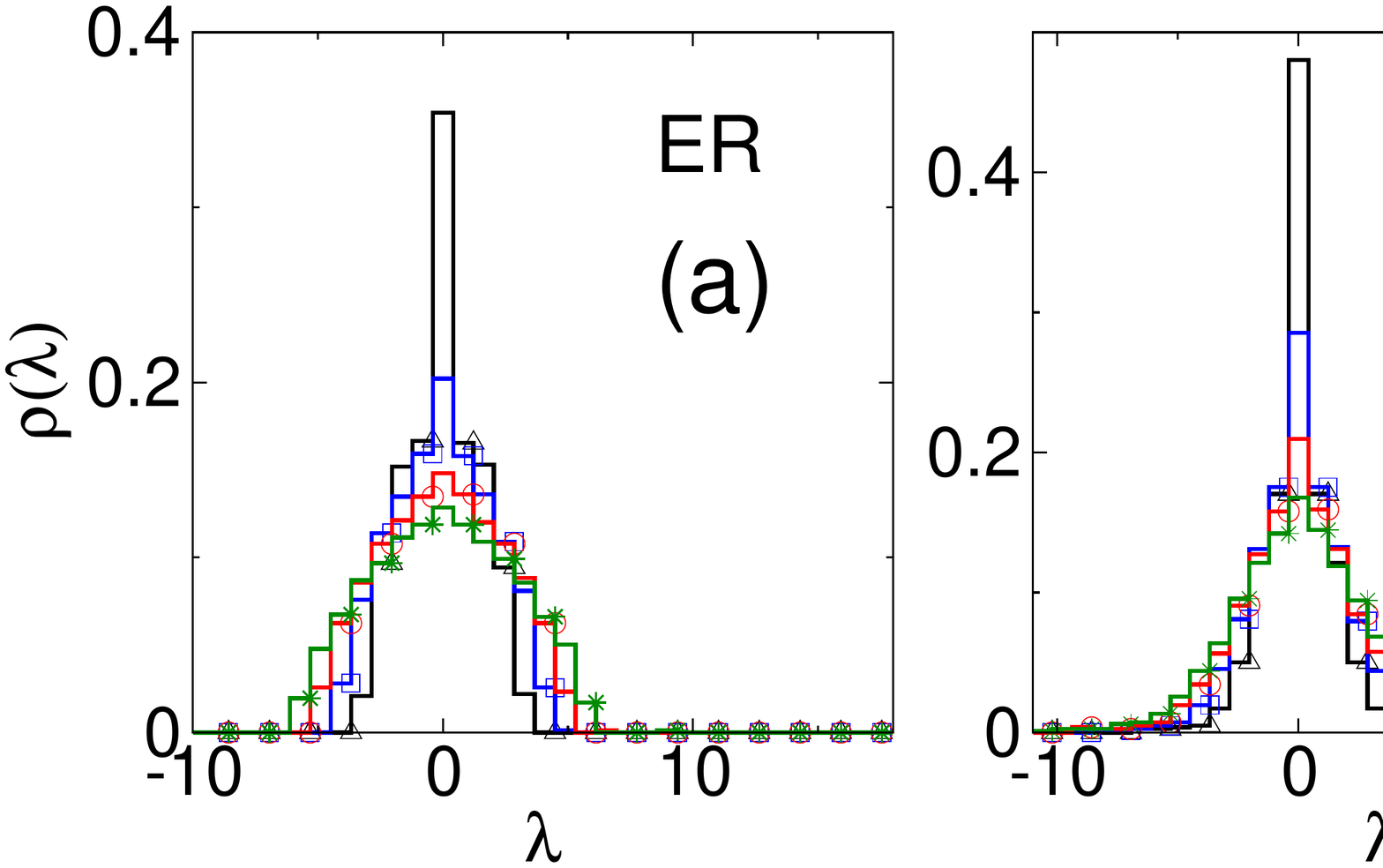}}
\caption{The spectral density of ER random networks and scale-free networks for different average degrees and $N=1000$. {\color{black} $\vartriangle$}, {\color{blue} $\square$}, {\color{red} $\circ$} and {\color{green} $\ast$} represent the data points of density distribution for $\langle k \rangle$=2, 4, 6 and 8, respectively. Reproduced figure (Figure 2) from A. Yadav and S. Jalan, Chaos 25, 043110 (2015), with the permission of AIP Publishing. DOI: \href{https://aip.scitation.org/doi/10.1063/1.4917286}{10.1063/1.4917286}.}
\label{Spectral_density_duplication}
\end{figure}

Further, it was found that the density distribution at very low average degree yields a peak at the zero eigenvalue, while with an increase in $\langle k \rangle$, the peak of the density distribution flattened (Fig.~\ref{Spectral_density_duplication} (a)). This result was true in case of scale-free networks as well, indicating that low average degree favors duplication. This can be explained in terms of the possible number of ways of duplication which is the ratio of the possible number of combinations of duplication possessed by two $k$-degree nodes to the possible number of combinations of random connections of those nodes. This is given as $\frac{k\,!}{N^{k}}$, where $N$ is the total number of nodes. As $k$ increases, the possible number of ways of duplication drastically decreases, thus explaining why low degree supports duplication. Moreover, by virtue of preferential attachment property, the low degree nodes have the highest probability to connect with the hubs of the network, which increases the likelihood of any two nodes to have the same neighbors, leading to a pair of duplicate nodes. Although even with increase in average degree, the density distribution remains triangular, there is flattening of the peak (Fig.~\ref{Spectral_density_duplication} (b)). This might be because 
the low degree nodes also tend to acquire connections with nodes other than the hubs. In order to check whether preferential attachment turns out to be one of the factors implicating in degeneracy, results were presented for configuration model networks constructed using the power-law degree sequence of the scale-free network as input but connections drawn randomly. In spite of having randomly assigned connections, they display a much higher zero degeneracy as compared to the ER random networks, indicating that it is the particular (the power-law) degree sequence emerges as a probable reason behind high degeneracy at zero eigenvalue.

Although node duplication provides a clue to the origin of zero degeneracy \cite{Jost}, there could be impact of network architecture on the duplication phenomenon. As discussed, for scale-free networks generated using preferential attachment mechanism, it fails to provide a quantitative measure of actual degeneracy 
observed in real world networks \cite{Dorogovtsev_PRE_2003}, indicating the contribution from other factors. Scale-free behaviour or sparseness of real world networks have been 
argued out to be other reasons responsible for degeneracy at the zero eigenvalues \cite{Dorogovtsev_PRE_2003,Aguiar_2005,SJ_PhysicaA_2014}.
\begin{figure}
\centerline{\includegraphics[width=0.5\columnwidth]{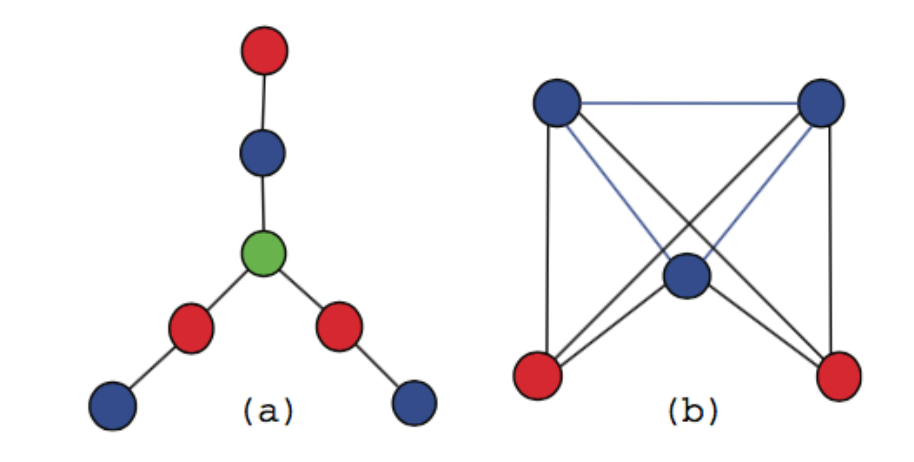}}
\caption{(a) and (b) have the same number of $\lambda_{-1}$ but different structures. The $\lambda_{-1}$ degeneracy of (a) occurs through the condition iii), whereas for (b) it results from the condition ii). Reproduced figure (Figure 2) with permission from L. Marrec and S. Jalan, Analysing degeneracies in networks spectra, EPL 117 48001 (2017). DOI: \href{http://iopscience.iop.org/article/10.1209/0295-5075/117/48001/meta}{10.1209/0295-5075/117/48001}.}
\label{Loic1}
\end{figure}

In addition to $0$ eigenvalue, there are reports on occurrence of $-1$ and $-2$ eigenvalues \cite{Meighem_book_2010}. Few papers related $0$ and $-1$ eigenvalues to stars and cliques, respectively \cite{Dorogovtsev_PRE_2003,Goh_PRE_2001,Kamp_PRE_2005}. However, it was observed that graphs in the absence of stars and cliques can still show a degeneracy at $0$ and $-1$ eigenvalues. As a result, these reasons are not exhaustive and it turns out that the origins of degeneracy at these eigenvalues are more complex. Degeneracy at $-1$ eigenvalue in networks spectra are known to be associated with some typical structures. However, the study of eigenvalues and their multiplicities is not sufficient to determine the number and size of these structures in networks (Fig.~\ref{Loic1}). It was found that eigenvectors associated with the degenerate eigenvalues shed light on the structures contributing to the degeneracy \cite{SJ_Loic} illustrating the nodes that contribute to degenerate eigenvalues. 

\section{Conclusion and Other Directions}
In this review, we present a comprehensive overview of the characteristic properties and the importance of eigenvalues spanning the entire range of the spectrum of adjacency matrices drawn on networks. Based on the intrinsic properties of the eigenvalues and the phenomena they capture, we segregate the spectra of networks under three major divisions: extremal eigenvalues, the bulk of the eigenvalues and degenerate eigenvalues. Under each major division, we put forward an account of most of the pioneering works that have enabled us to strengthen our understanding of spectra of networks. A detailed look at the spectra of networks reveals how the bounds of eigenvalues change depending on the structural attributes of networks, for example average degree, largest degree, degree distribution, degree-degree correlation, etc. with particular emphasis on the popular model networks, namely 1-d lattice, ER random networks, scale-free networks and small-world networks. These bounds are important for understanding various phenomena occurring in nature, for instance, epidemic spread, synchronization of coupled oscillators, the stability of couplings architecture in the brain, etc. We have also discussed the rich statistics that the eigenvalues yield as well as the implications of eigenvalues in understanding the real-world systems represented using networks.  

Along with the eigenvalues, the associated eigenvectors of a graph are also intimately related to important topological features such as the diameter, the number of cycles, and the connectivity properties of the graph. Understanding the localization properties of eigenvectors has helped to render a better understanding of the disease-spreading phenomenon in underlying networks \cite{Dorogovtsev_PRL_2012}. Recent investigations on eigenvector localization using inverse participation ratio have revealed the collective influence of a set of distinct structural as well as spectral features on the localization properties of the principal eigenvector \cite{SJ_Prio_2017}.

Further, in addition to the spectra of adjacency matrices, eigenvalues of Laplacian matrices, particularly the largest and the first non-zero, have been
investigated in great details as they provide information about the stability and convergence of
diffusion processes and the synchronized state on networks \cite{Chung_book,Sitabhra}.
In their celebrated work, Pecora and Carroll have shown how stability of synchronized state of diffusively coupled chaotic dynamical units on networks can be easily determined by Laplacian eigenvalue ratio \cite{Pecora_PRL_1998}.
Primarily driven by this discovery, there had been a spurt in activities for analyzing Laplacian eigenvalue ratio of networks having various different structural characteristics \cite{Sitabhra}. Further, in diffusion processes, the relaxation rate has been shown to be governed by the corresponding eigenvalues of the normalized 
Laplacian \cite{Motter_PRE_2005}.
Laplacian spectra have also been used to understand synchronization of weighted networks, where weights have been shown to play a pivotal role in synchronizability of random networks \cite{Weighted_Kurths}. It has also been shown that introduction of weights enhances the complete synchronization of identical dynamical units in scale-free networks \cite{Weighted_Boccaletti}.
Furthermore, recently it has been realised that
many real-world complex systems having different types of interactions can be modeled using multiplex or multilayer networks framework \cite{kurths_net_net,Boccaletti_PhysRep_2014,SJ_CS_EPL_2016}. Motivated from the success of this new framework in predicting various behaviours of interacting units, spectra of adjacency and Laplacian matrices have also been investigated, particularly in the context of understanding occurrence of extreme events \cite{SJ_SG_EPL_2016} and diffusion processes \cite{Arenas_PRL_2013} on multilayer networks. Using Laplacian eigenvalue ratio, it has been shown that optimization complexity of a system having multiple layers of interactions can
be drastically reduced and synchronization on the entire multilayer network can be regulated by one of
its layer \cite{SJ_SKD_2017}.

%--------------------------------------------------------------

%----------------------------------------------------------------

%-------------------------------------------------------------------------------------------------------------------------

\section{Acknowledgements}
SJ acknowledges support by grant of DST, Govt. of India (EMR/2016/001921), BRNS project (37(3)/14/11/2018-BRNS/37131) and 
Ministry of Education and Science of the Russian Federation Agreement No. 074-02-2018-330.
We are thankful to the complex systems lab members, particularly, Priodyuti Pradhan, Saptarshi Ghosh,
Alok Yadav, Pramod Shinde and Ajay Deep Kachhvah for simulating discussions and their inputs on this review article.

\end{document}